\documentclass[citeautoscript,floatfix,aps,prb,twocolumn,superscriptaddress]{revtex4-2}
\usepackage{graphicx}
\usepackage{dcolumn}
\usepackage{bm}
\usepackage{amssymb}
\usepackage[normalem]{ulem} 
\usepackage{nicefrac}
\usepackage{hyperref}
\usepackage{booktabs}
\usepackage{import}
\usepackage{makecell}
\usepackage[caption=false]{subfig}
\usepackage{graphicx}
\usepackage{amsmath}
\usepackage{amssymb}
\usepackage{dcolumn}
\usepackage[version=4]{mhchem}
\usepackage{latexsym}
\usepackage{rotating}
\usepackage{epstopdf}
\usepackage[usenames,dvipsnames]{xcolor}
\usepackage{float}
\usepackage{soul}
\usepackage{epsfig}
\usepackage{psfrag}
\usepackage{natbib}
\usepackage{eucal}
\usepackage{mathrsfs}
\usepackage{braket}
\usepackage{enumerate}
\usepackage{longtable}
\usepackage{hyperref}
\usepackage{amsfonts}
\usepackage{dcolumn} 
\usepackage{changes}
\usepackage{appendix}
\usepackage{mathrsfs}
\usepackage{tcolorbox}
\usepackage{tabularx}
\usepackage{booktabs,tabularx,makecell,array}
\usepackage{color} 
\usepackage{hyperref}
\usepackage[OT1]{fontenc}

\newcolumntype{Y}{>{\raggedright\arraybackslash}X} 

\hypersetup{
	colorlinks=true,final=true,
	linkcolor=blue,
	citecolor=blue,
	filecolor=blue,
	urlcolor=blue,
}

\graphicspath{ {./images/} }

\begin{document}
	
	\title{Chiral and Topological Phonons in the Square-Octagon Lattice}
	\author{Ravi Kiran}
	\email{ravieroy123@iitkgp.ac.in}
	\affiliation{Department of Physics, Indian Institute of Technology Kharagpur, Kharagpur 721302, West Bengal, India}
	
	\author{A. Taraphder}
	\email{arghya@phy.iitkgp.ac.in}
	\affiliation{Department of Physics, Indian Institute of Technology Kharagpur, Kharagpur 721302, West Bengal, India}
	
	\begin{abstract}
		Controlling chirality and topology in phonon systems requires microscopic routes to reshape optical modes, open symmetry-governed gaps, and understand the robustness of the resulting chiral vibrational structure against disorder. The square-octagon lattice provides a natural platform for this purpose because its four-site basis generates internal optical modes and symmetry-allowed band crossings that are sensitive to the range of elastic couplings. We show that extended interactions, mass imbalance, and time-reversal symmetry (TRS) breaking provide an effective means to control these features. In particular, third-nearest-neighbor coupling releases otherwise nearly frozen optical branches, demonstrating that longer-range restoring pathways are essential for activating internal motions of the multi-site basis. Once TRS is broken, the symmetry-allowed line crossings are converted into avoided crossings, producing phonon modes with finite circular polarization. The resulting Berry curvature is concentrated near these gapped regions, and the relevant bands acquire opposite Chern numbers, identifying a topological chiral-phonon phase. We further find that the chiral sector remains most robust under mass disorder and is more strongly suppressed by spring disorder, while local chiral motion and the spectral identity of the chiral sector survive more strongly than the coherent global chiral response.
	\end{abstract}
	
	\maketitle
	
	\section{Introduction}
	
	Phonons are fundamental collective excitations \cite{tamm1930quantentheorie} of crystalline solids and play a central role in determining thermal, mechanical, and transport properties. In recent years, considerable attention has been given to phonon modes that carry angular momentum, because they provide a route for coupling lattice motion to electronic, magnetic, and optical degrees of freedom. Among these, chiral phonons are of particular interest. Originally proposed by Zhang \textit{et al.}~\cite{zhang2015chiral} and later observed experimentally~\cite{zhu2018observation,ueda2023chiral,pols2025chiral}, these modes exhibit circularly polarized atomic motion and finite angular momentum, in contrast to conventional linearly polarized vibrations. Their unusual rotational character has opened new possibilities for understanding and controlling dynamical phenomena in quantum materials and engineered mechanical systems~\cite{luo2023large,sasaki2021magnetization,kim2023chiral,saito2019berry,qin2012berry,zhang2025weyl,zhang2023weyl}.
	
	Recent theoretical work \cite{juraschek2025chiral} has further clarified that angular-momentum carrying modes constitute a broader class, often collectively called \emph{axial phonons}, whereas only those modes that further break improper rotational symmetries are \emph{chiral}. This more refined view offers a logically consistent symmetry-based basis for interpreting their physical consequences and helps clarify some ambiguities that have arisen in the literature regarding what should properly be called a chiral phonon. The emergence of chiral phonons primarily requires the breaking of time-reversal symmetry (TRS) and/or inversion symmetry. When both TRS and inversion symmetry are broken, truly chiral phonons can emerge with well-defined handedness \cite{reddy2024coexistent,mishra2025chiral,ishito2023truly,ohe2024chirality}.
	
	The interplay between symmetry, chirality, and topology has therefore become an important theme in lattice dynamics. The geometry and symmetry of the underlying lattice play a central role in determining the nature of phonon excitations. In hexagonal lattices, for instance, the threefold rotational symmetry at high-symmetry points gives rise to phonon eigenmodes carrying quantized pseudoangular momentum \cite{zhang2015chiral, zhu2018observation}. Beyond hexagonal systems, chiral phonons have also been predicted in square lattices, where their emergence requires explicit breaking of TRS \cite{wang2022chiral}. In the kagome lattice, chiral phonons with pronounced circular polarization have been observed at high-symmetry points; unlike in the hexagonal case, all three sublattices there vibrate with the same handedness \cite{chen2019chiral}. Triangular lattices have further been shown to host chiral phonons with nontrivial topological character, where the chirality remains robust against variations in sublattice mass and interlayer coupling, reflecting intrinsic topological protection \cite{xu2018topological}. These studies show that multi-sublattice structures with symmetry-protected crossings provide natural platforms for the emergence of circular polarization, Berry curvature, and band topology. An interesting question, however, is how robust these features remain once one moves beyond the ideal periodic limit. Disorder can substantially modify lattice vibrations through mode broadening, mixing, and localization \cite{monthus2010anderson, allen1993thermal, seyf2017rethinking}.
	
	This issue is especially important because real materials and phononic metamaterials inevitably contain disorder in the form of mass inhomogeneity, bond-strength variation, defects, or fabrication-induced irregularities. For chiral phonons the handedness of a mode is encoded in the relative phase and amplitude of atomic motion across the sublattices and disorder can mix nearby modes, broaden spectral features, and modify or even suppress the net chirality. From the perspective of topological lattice dynamics, disorder also provides a stringent test of whether the chiral response is a fragile consequence of perfect translational symmetry or a more robust feature that survives in realistic settings. A natural framework for addressing both the clean and disordered problems on equal footing is the dynamical-matrix method. In the translationally invariant case, it yields the phonon bands and polarization structure in momentum space and in the presence of disorder, it remains applicable in real space even when crystal momentum is no longer a good quantum number.

	Alongside these developments, artificial acoustic and phononic structures have emerged as powerful platforms for realizing and probing novel vibrational phenomena~\cite{huber2016topological, liu2007phononic, croenne2011band, gantzounis2011multiple,muhlestein2020effective,yang2014homogenization}. Studies of phononic crystals and periodic elastic composites have demonstrated band gaps, unusual dispersion relations, and strong control over wave propagation~\cite{nikitenkova2010dispersion,kushwaha1993acoustic,tanaka2000band,mei2005multiple,vasseur2001experimental,yang2004focusing,wilm2003out,liu2000locally,sigalas2000theoretical,lambin2001stopping,kafesaki1999multiple}. More recently with the birth of the field of electronic topological insulators, topology also found a new path back to classical systems \cite{kane2014topological, prodan2009topological, berg2011topological, susstrunk2016classification}. These advances motivate the search for lattice geometries in which chirality and topology can be realized and controlled in a simple setting.
	
	In this context, the square--octagon lattice provides an interesting and relatively unexplored setting for studying vibrational chirality. Because of its four-site unit cell and nontrivial connectivity, the lattice supports a rich phonon spectrum with symmetry-allowed crossings, internal optical modes, and strong mode hybridization. In electronic systems, the same geometry is known to give rise to flat bands and van Hove singularities (VHS)~\cite{pal2018nontrivial,oriekhov2021orbital}. The significance of the present work is to show that these geometrical features can also organize chiral and topological phonon responses. In particular, the square--octagon geometry allows us to connect internal optical-mode activation, longer-range elastic connectivity, and disorder-induced changes in local and global chirality within a single phonon model.
	
	In this work, we study the phonon spectrum of the square--octagon lattice within a classical spring-mass model using a dynamical-matrix formalism. We characterize the clean-band structure, identify symmetry-allowed Dirac crossings along $\Gamma$--$M$ and $\Gamma$--$X$, and analyze the resulting topological phonon bands using Berry curvature and Chern numbers. We then examine the effect of disorder on the chirality of these modes and assess how local circular motion, global circular polarization, and topological character are affected in the absence of perfect periodicity.
	
	The remainder of this paper is organized as follows. In Sec.~\ref{sec:theoretical_background}, we introduce the model and outline the theoretical framework. Section~\ref{sec:results} presents and discusses the numerical results. Finally, Sec.~\ref{sec:summary_conclusion} summarizes the main findings and provides concluding remarks.

	\section{Model and Methodology} 
	\label{sec:theoretical_background}
	\subsection{General framework}
	\label{subsec:general_framework}
	
	Consider a $d$-dimensional Bravais lattice with primitive vectors
	$a=(a_1,\dots,a_d)$ and a basis of $s$ atoms at equilibrium positions
	$\{\tau_i\}_{i=1}^s$ within each unit cell. A site is labeled by $(R,i)$,
	with Cartesian position
	\begin{equation}
		\mathbf r_{R,i}=R+\tau_i
		\qquad
		R=\sum_{\mu} n_\mu a_\mu
	\end{equation}
	Let $\mathbf u_{R,i}(t)\in\mathbb R^d$ denote the displacement of basis atom
	$i$ in cell $R$. The general linear equations of motion under harmonic approximation can be written as,
	
	\begin{equation}
		\label{eq:eom_real}
		\begin{aligned}
			\sum_{R',j}
			\Big[ \,
			&M_{ij}\delta_{RR'}\,\partial_t^2
			+\mathcal A_{ij}(R-R')\,\partial_t \\
			&+\mathcal K_{ij}(R-R')
			\Big]
			\mathbf u_{R',j}(t)=0
		\end{aligned}
	\end{equation}
	
	where $M_{ij}=m_i\delta_{ij}\,I_d$ is the mass matrix, $\mathcal K_{ij}(R)$ are the harmonic force-constant blocks, and $\mathcal A_{ij}(R)$ is an effective velocity-dependent kernel representing a reactive gyroscopic or Lorentz-like contribution. Velocity-dependent phonon terms of this form arise in gyroscopic phononic crystals and mechanical metamaterials, and also in spin-phonon or phonon Hall models where magnetic degrees of freedom generate effective TRS-breaking forces on lattice vibrations ~\cite{wang2015topological,nash2015topological,sheng2006theory,qin2012berry,zhang2014angular}. A brief derivation of the corresponding velocity dependent contribution from an antisymmetric first order in time lattice Lagrangian is given in the Supplemental Material \cite{kiran2026supplementary}.
	
	Using the Bloch convention,
	 
	$$
	\mathbf u_{R,i}(t)=\boldsymbol{\epsilon}_i(\mathbf k)e^{i\mathbf k\cdot(R+\tau_i)}e^{-i\omega t},
	$$
	and defining the \(N(=ds)\)-component vector $\mathbf u(\mathbf k)=\big(\boldsymbol{\epsilon}_1(\mathbf k),\dots,\boldsymbol{\epsilon}_s(\mathbf k)\big)^T,$
	one obtains the frequency-domain quadratic eigenvalue problem
	
	\begin{equation}
		\label{eq:qep_u}
		\Big[
		K(\mathbf k)-\omega^2 M-i\omega A(\mathbf k)
		\Big]\mathbf u(\mathbf k)=0
	\end{equation}
	The Fourier-space matrices are
	\begin{align}
		\label{eq:KA_fourier}
		K_{ij}(\mathbf k)
		&=
		\sum_R \mathcal K_{ij}(R)\,
		e^{-i\mathbf k\cdot R}\,
		e^{i\mathbf k\cdot(\tau_j-\tau_i)}
		\nonumber\\
		A_{ij}(\mathbf k)
		&=
		\sum_R \mathcal A_{ij}(R)\,
		e^{-i\mathbf k\cdot R}\,
		e^{i\mathbf k\cdot(\tau_j-\tau_i)}
	\end{align}
	
	Introducing the mass-normalized coordinates $\mathbf v(\mathbf k)=M^{1/2}\mathbf u(\mathbf k)$ Eq.~\eqref{eq:qep_u} becomes
	
	\begin{equation}
		\label{eq:qep_v}
		\Big[
		D_0(\mathbf k)-\omega^2 I
		-i\omega \widetilde A(\mathbf k)
		\Big]\mathbf v(\mathbf k)=0
	\end{equation}
	where
	\begin{equation*}
		\label{eq:D0_Atilde}
		\begin{aligned}
			D_0(\mathbf k)
			&=M^{-1/2}K(\mathbf k)M^{-1/2}\\
			\widetilde A(\mathbf k)
			&=M^{-1/2}A(\mathbf k)M^{-1/2}
		\end{aligned}
	\end{equation*}
	For convenience we introduce the Hermitian kernel $G(\mathbf k)\equiv -\,i\widetilde A(\mathbf k)$ in terms of which Eq.~\eqref{eq:qep_v} may be written as
	\begin{equation}
		\label{eq:qep_v_G}
		\Big[
		D_0(\mathbf k)-\omega^2 I
		+\omega G(\mathbf k)
		\Big]\mathbf v(\mathbf k)=0
	\end{equation}
	
	At the level of the present effective theory, the kernel $G(\mathbf{k})$ is interpreted as arising from magnetoelastic coupling between lattice displacements and an underlying gapped magnetic degree of freedom. It is useful to separate its overall strength from its lattice-resolved structure by writing $G(\mathbf{k}) = g\,\Gamma(\mathbf{k})$, where $\Gamma(\mathbf{k})$ is a dimensionless bond-resolved kernel fixed by lattice geometry and symmetry, while $g$ controls the overall magnitude of the induced interaction. In the present effective model, we parameterize the scale as
	\begin{equation}
		\label{eq:geff_main}
		g \sim \frac{\lambda^2 M_z}{\Omega_m},
	\end{equation}
	where the proportionality constant depends on microscopic details of the magnetic susceptibility and magnetoelastic coupling. \(\lambda\) denotes the magnetoelastic coupling, $M_z$ is the out-of-plane time-reversal odd magnetic polarization, and $\Omega_m$ is the characteristic energy of the gapped magnetic mode. This effective kernel lifts the degeneracies of the conservative Dirac crossings, opens gaps, and endows the resulting phonon modes with finite chirality. A motivation for the phenomenological choice of the TRS-breaking scale $g$ is provided in Supplemental Material \cite{kiran2026supplementary}.
	
	\subsection{Phonon density of states}
	\label{subsec:sqoct_dos}
	
	In the following section, we analyze both the phonon band structure and the corresponding density of states(DOS). The total phonon DOS is calculated from the phonon frequencies over the full Brillouin zone (BZ) as,
	
	\begin{equation}
		D(\omega)
		=
		\frac{1}{N_k}
		\sum_{\mathbf k,n}
		\delta(\omega-\omega_n(\mathbf k)),
	\end{equation}
	where $N_k$ is the total number of sampled momenta and $n$ labels the phonon branches. In the numerical calculation, the delta function is replaced by a normalized Gaussian broadening,
	\begin{equation}
		\delta(\omega-\omega_n)
		\rightarrow
		\frac{1}{\sigma\sqrt{2\pi}}
		\exp\left[
		-\frac{(\omega-\omega_n)^2}{2\sigma^2}
		\right],
	\end{equation}
	so that
	\begin{equation}
		D(\omega)
		=
		\frac{1}{N_k}
		\sum_{\mathbf k,n}
		\frac{1}{\sigma\sqrt{2\pi}}
		\exp\left[
		-\frac{(\omega-\omega_n(\mathbf k))^2}{2\sigma^2}
		\right].
	\end{equation}
	
	\begin{figure}
		\includegraphics[width=\linewidth]{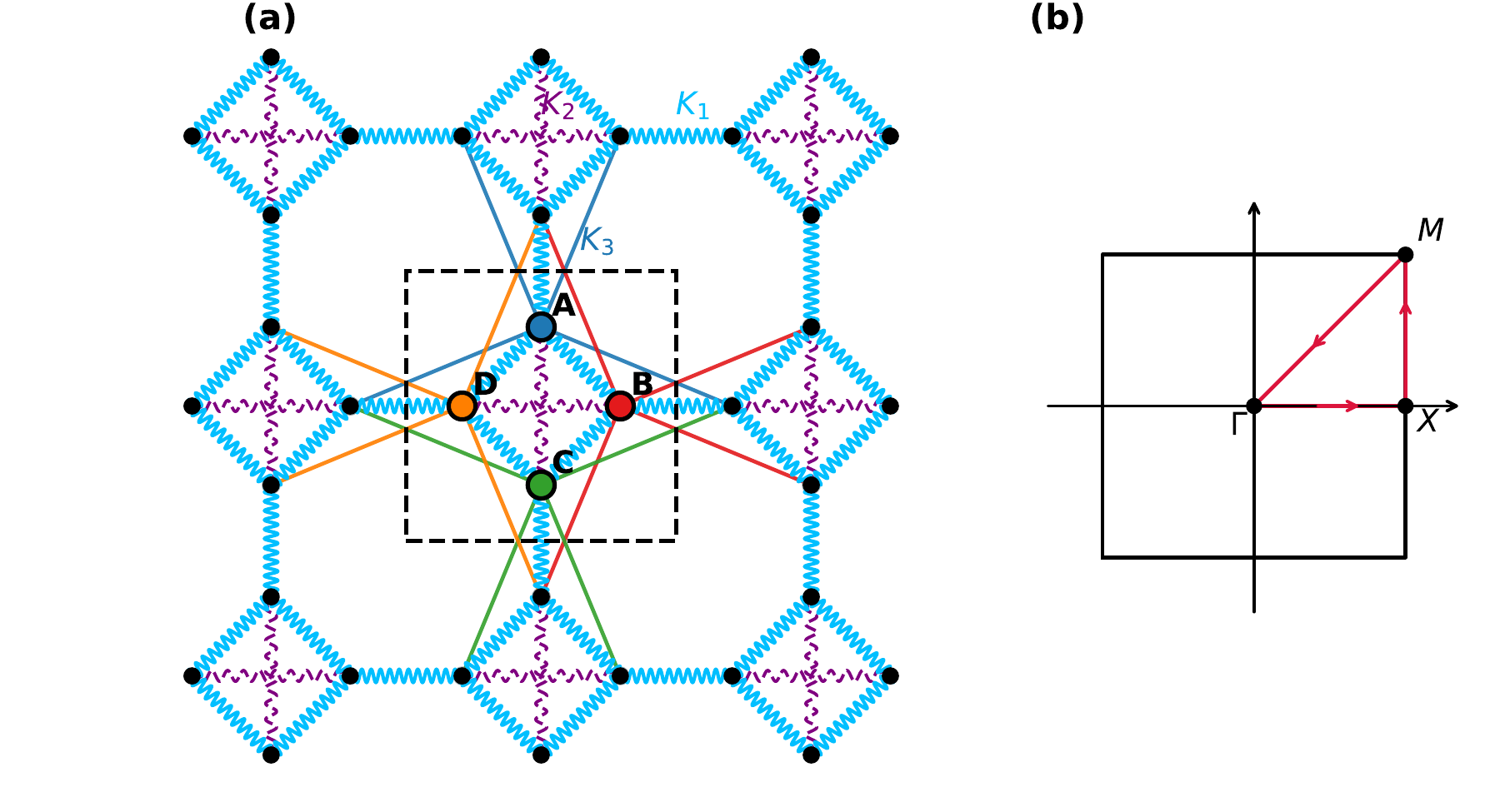}
		\caption{(a) Schematic representation of a 2D square-octagon lattice model, with a repeating unit cell outlined by a black dashed square containing four atomic sites. Nearest-neighbor (NN) springs are shown as solid blue sinusoidal lines, next-nearest-neighbor (NNN) springs as dashed purple sinusoidal lines, and third-nearest-neighbor connections as straight colored bonds extending from the central cell. The coupling constants $K_1$, $K_2$, and $K_3$ correspond to representative NN, NNN, and third-NN interactions, respectively. (b) Brillouin zone showing the high-symmetry points \(\Gamma\), \(X\), and \(M\), together with the \(\Gamma\)--\(X\)--\(M\)--\(\Gamma\) path used for the phonon-band calculations.}
		\label{fig:so_lattice}
	\end{figure}
	
	\subsection{Square-octagon lattice up to third nearest neighbors}
	\label{subsec:sqoct_main}
	
	For the present work we consider oscillations of a square-octagon lattice (Fig.~\ref{fig:so_lattice}) with particles of masses $m_\alpha$ ($\alpha \in \{A,B,C,D\}$) positioned at the basis sites of the lattice. In equilibrium, the particle centers coincide with these basis positions, separated by a characteristic distance $a$. Each particle interacts with its nearest-neighbor (NN), next-nearest-neighbor (NNN), and third-nearest-neighbor (3rd-NN) through springs of stiffness constants $K_1$, $K_2$, and $K_3$, respectively. Explicit bond vectors, Bloch phases, and block contributions for the square-octagon geometry are provided in the Supplemental Material \cite{kiran2026supplementary}.
	
	\section{Results}
	\label{sec:results}
	\subsection{Phonon band structure and DOS}
	\label{sec:phonon_bandstructure_and_dos}

	\begin{figure}
		\centering
		\includegraphics[width=\linewidth]{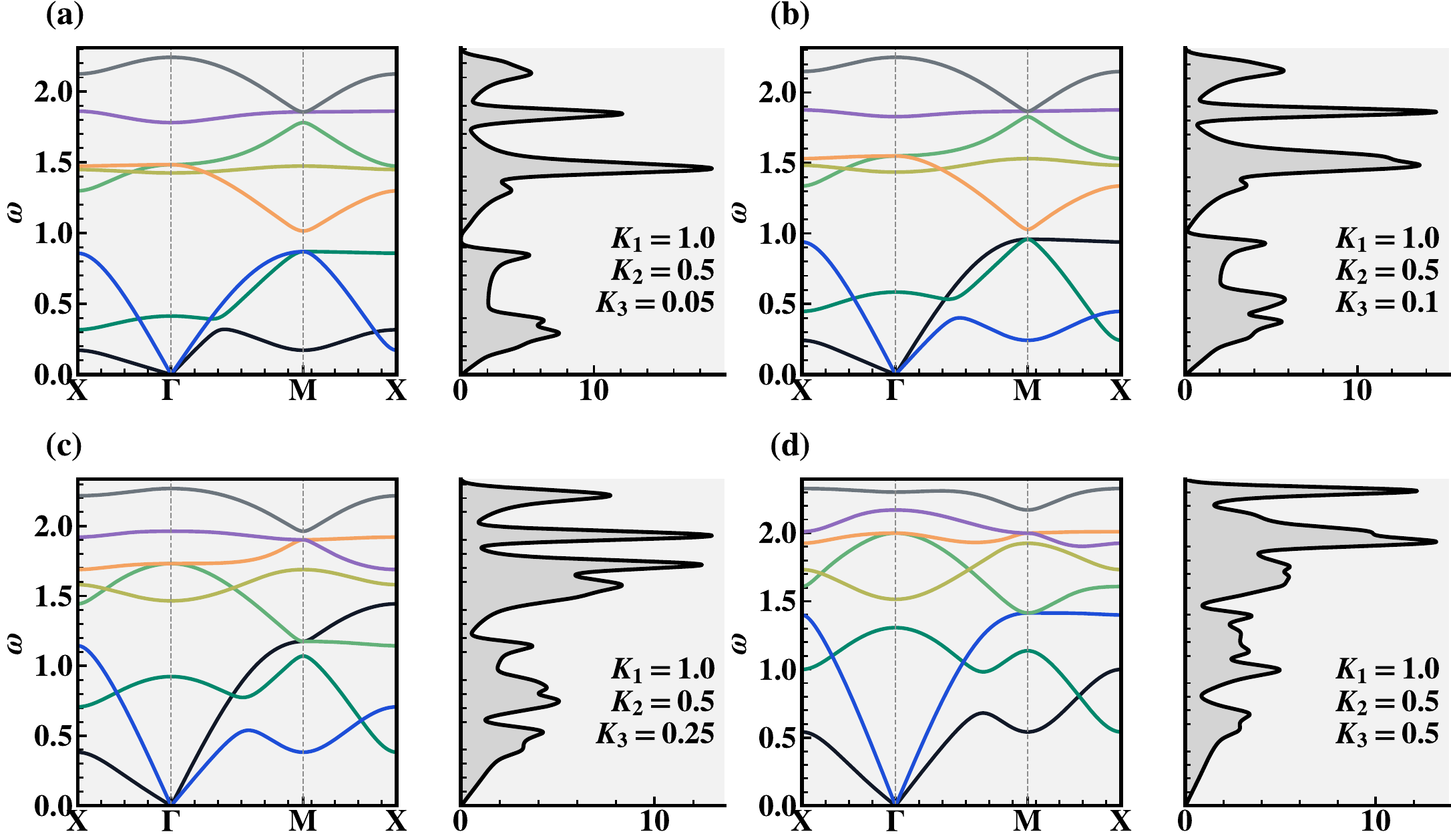}%
		\caption[Phonon dispersions and DOS for increasing third-neighbor coupling]
		{
			Phonon dispersions along the $X\!-\!\Gamma\!-\!M\!-\!X$ path and the corresponding DOS for the uniform mass case ($m_\alpha = 1$), with fixed NNN coupling $K_2 = 0.5$ and increasing third-neighbor coupling: (a) $K_3 = 0.05$, (b) $K_3 = 0.1$, (c) $K_3 = 0.25$, and (d) $K_3 = 0.5$. There are $2\times4=8$ phonon branches. The two branches that vanish at $\Gamma$ are the acoustic branches, corresponding to rigid translations of the lattice, while the remaining six finite-frequency branches. The two acoustic branches are shown using darker colors(blue and black) to distinguish them from the optical sector.
		}
		\label{fig:K3_uniform_masses}
	\end{figure}
 
	We first consider the uniform mass lattice, $m_\alpha=1$ for all sublattices, and examine the effect of varying the third-NN coupling $K_3$ at fixed $K_1=1$ and $K_2=0.5$, as shown in Fig.~\ref{fig:K3_uniform_masses}. For a $d$-dimensional lattice with $s$ atoms in the basis, the phonon spectrum contains $ds$ branches. In a translationally invariant lattice, the acoustic branches are identified as the $d$ modes whose frequencies vanish at the $\Gamma$ point, corresponding to rigid translations of the lattice. The remaining $d(s-1)$ branches have finite frequency at $\Gamma$ and are referred to as optical branches. For the present 2D square-octagon lattice with four sites per unit cell, there are eight phonon branches, two acoustic branches and six optical branches.
	
	The inclusion of $K_3$ is essential in this model, since in its absence several optical branches remain nearly dispersionless. The corresponding $K_3=0$ results are presented separately in the Supplemental Material~\cite{kiran2026supplementary}. Once $K_3$ is turned on, these nearly frozen optical modes acquire dispersion and the optical sector is progressively reorganized, while the acoustic behavior near $\Gamma$ remains intact.
	
	For weak third-neighbor coupling, Fig.~\ref{fig:K3_uniform_masses}(a), the effect of $K_3$ is still modest: the optical branches retain a relatively simple structure and the DOS shows a few well-defined van Hove features. As $K_3$ increases through Figs.~\ref{fig:K3_uniform_masses}(b) and \ref{fig:K3_uniform_masses}(c), the middle and upper optical branches shift and change curvature, leading to a redistribution of spectral weight and a richer DOS structure. When $K_3$ becomes comparable to $K_2$, Fig.~\ref{fig:K3_uniform_masses}(d), the higher-frequency optical modes are substantially reconstructed, with stronger hybridization among optical branches and a broader spread of DOS features.

	\begin{figure}
		\centering
		\includegraphics[width=\linewidth]{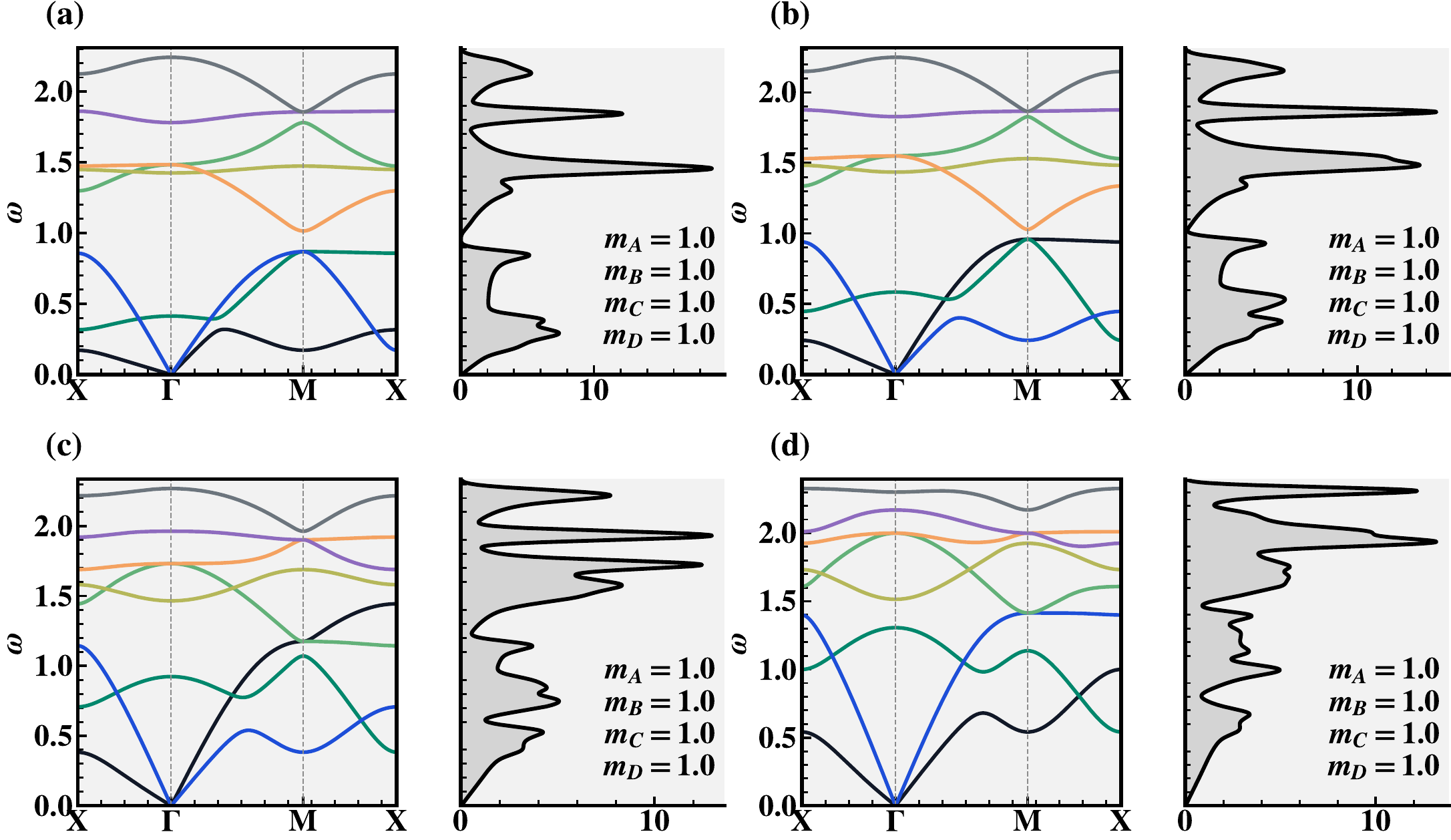}%
		\caption{Phonon dispersions along the $X$--$\Gamma$--$M$--$X$ path and the corresponding DOS for the square-octagon lattice with fixed couplings $K_1=1.0$, $K_2=0.5$, and $K_3=0.1$. (a-c) With increasing contrast in the masses of sublattices B and D, there occurs a decrease in the frequency of certain optical branches, leading to the growth of DOS and greater spectral distinction of phonon branches. For large values of the mass contrast, quasi-flat optical branches emerge, implying that vibrations become localized on heavy sublattices. (d) For the light impurity case, some frequencies move upward, forming an isolated band at higher energies and a sharp peak in DOS.}
		\label{fig:K3_mass_variations}
	\end{figure}
	 
	For fixed couplings $K_1=1$, $K_2=0.5$, and $K_3=0.1$, we next examine how mass contrast reshapes the phonon dispersions and the total DOS, as shown in Fig.~\ref{fig:K3_mass_variations}. Beginning from the weak-contrast case in Fig.~\ref{fig:K3_mass_variations}(a), the main effect of unequal masses is to lift the band degeneracies present in the corresponding uniform-mass lattice and to reorganize the optical branches into a more clearly split set of modes. As the contrast is increased in Fig.~\ref{fig:K3_mass_variations}(b) and becomes more pronounced in Fig.~\ref{fig:K3_mass_variations}(c), the spectrum progressively separates into distinct frequency windows, while several branches become increasingly flat. This flattening is reflected directly in the DOS through the growth of sharper peaks, indicating an accumulation of vibrational states over narrow frequency intervals.
	
	In the large contrast regime, the heavy sublattices behave as comparatively slow oscillators, so that the vibrational dynamics become increasingly partitioned between light and heavy degrees of freedom. The resulting suppression of dispersion is accompanied by strong spectral crowding and enhanced vHS, providing a clear route toward phonon-band engineering through mass imbalance. A qualitatively different behavior emerges in Fig.~\ref{fig:K3_mass_variations}(d), where a light impurity is introduced on the $B$ sublattice. In this case, the most striking effect is the appearance of high-frequency modes above the main band structure. These modes give rise to pronounced peaks in the DOS and are consistent with impurity-dominated vibrations that are only weakly hybridized with the rest of the lattice.
	 
	 \subsection{Chiral Phonons}
	 \label{sec:chiral_phonons}
	 
	A recent study by Juraschek \textit{et al}.~\cite{juraschek2025chiral} introduced a systematic taxonomy for lattice vibrations that carry angular momentum. In 2D systems, the classification of chiral phonons differs fundamentally from the three-dimensional case because out-of-plane rotational degrees of freedom are absent. Consequently, 2D point-group operations do not include a mirror plane perpendicular to the angular-momentum axis. Without such an operation, the phonon modes cannot be mapped onto their mirror counterparts, leading to a breaking of improper rotational symmetry. These modes are therefore categorized as \emph{false-chiral}. Nevertheless, despite their false-chiral character, they still break the symmetry conditions required for the conventional ``chiral phonon'' label.
	 
	\begin{figure}
		\centering
		\includegraphics[width=\linewidth]{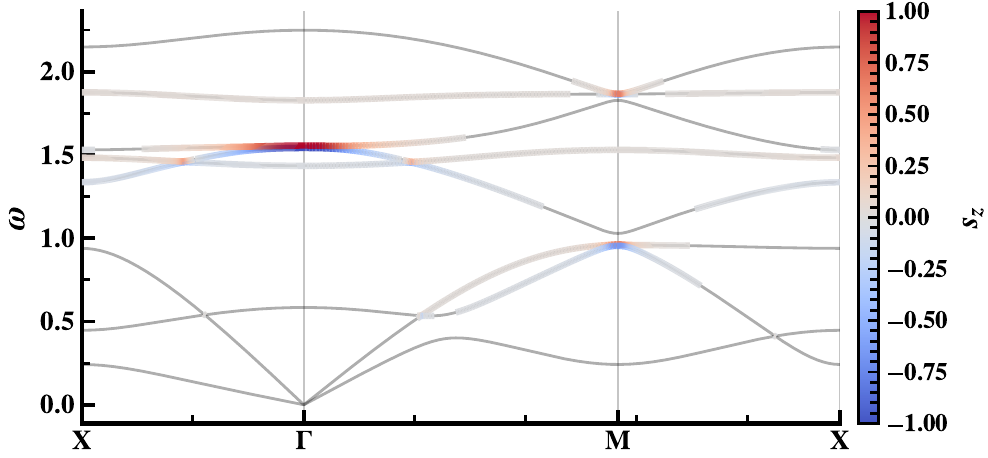}
		\caption{Phonon band structure of the square-octagon lattice in the presence of a finite TRS breaking term ($M_z=1.0$) for equal masses $(m_\alpha= 1$) and spring constants $K_1=1.0$, $K_2=0.5$, and $K_3=0.1$. The color scale shows the circular-polarization measure $s_z(\mathbf{k})$, with $s_z=\pm 1$ corresponding to perfectly chiral motion (opposite handedness) and $s_z=0$ to linear polarization. Finite $s_z$ concentrated near $\Gamma$ indicates the emergence of chiral phonon modes.}
		\label{fig:phonon_bands_Lz_uniform_masses}
	\end{figure} 
	
\begin{figure}
		\centering
		\includegraphics[width=\linewidth]{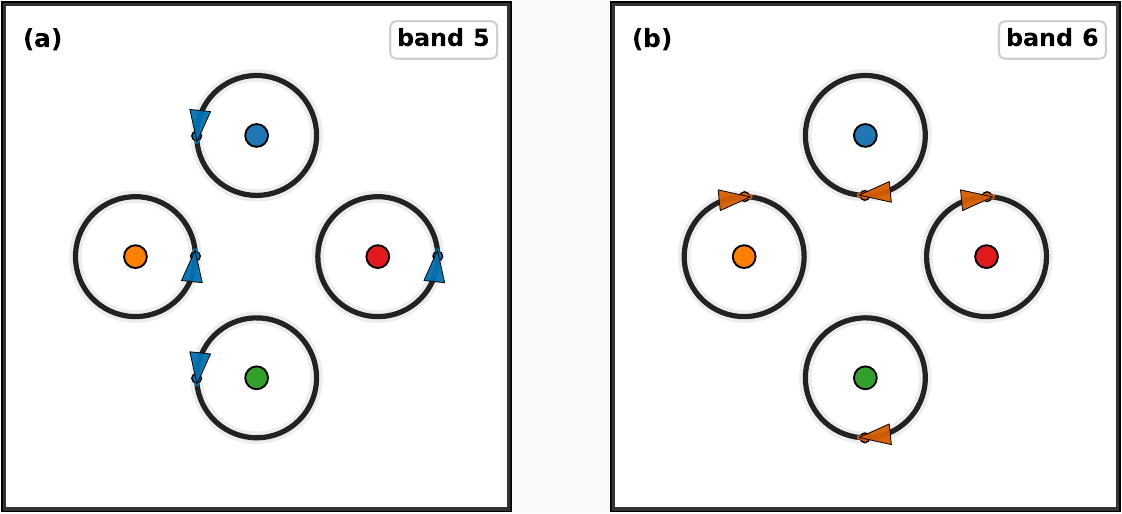}
		\caption{
			Phonon eigenmodes at the $\Gamma$ point for bands 5 and 6, shown for a system with uniform masses $m_\alpha=1.0$ and spring constants $K_1 = 1.0$, $K_2 = 0.5$, $K_3 = 0.1$ for $M_z = 1.0$. The blue and orange arrows denote counterclockwise and clockwise rotations, respectively, illustrating the alternating helicities of the chiral phonon modes.
		}
		\label{fig:chiral_phonon_modes_56}
	\end{figure}

	In Fig.~\ref{fig:phonon_bands_Lz_uniform_masses} we show the resulting phonon band structure for a system with uniform masses (\(m_\alpha = 1.0\)) and spring constants \(K_1 = 1.0\), \(K_2 = 0.5\), and \(K_3 = 0.1\), in the presence of a finite TRS breaking term (\(M_z = 1.0\)). For these parameters the introduction of a nonzero \(g\) lifts these degeneracies and leads to finite phonon angular momentum. To characterize the chirality of a phonon mode, we evaluate the circular polarization measure from the complex displacement eigenvector. For a phonon mode in band $n$ at momentum $\mathbf{k}$, let $u_{x,i,n}(\mathbf{k})$ and $u_{y,i,n}(\mathbf{k})$ denote the $x$ and $y$ components of the physical displacement of atom $i$ within the unit cell. The circular polarization of the mode is then quantified by
	
	\begin{equation}
		S_{z,n}(\mathbf{k})
		=
		\frac{
			2 \, \mathrm{Im} \, \left[
			\sum_i
			u_{x,i,n}^{*}(\mathbf{k})
			u_{y,i,n}(\mathbf{k})
			\right]
		}{
			\sum_i
			\left(
			\left|u_{x,i,n}(\mathbf{k})\right|^2
			+
			\left|u_{y,i,n}(\mathbf{k})\right|^2
			\right)
		}.
		\label{eq:sz_phonon}
	\end{equation}
	Here the displacement components are physical displacements, obtained from the mass-normalized eigenvectors by dividing the components on atom $i$ by $\sqrt{m_i}$. The quantity $S_{z,n}(\mathbf{k})$ lies in the interval $[-1,1]$. A value $S_{z,n}(\mathbf{k})=0$ corresponds to a linearly polarized mode, while $S_{z,n}(\mathbf{k})=\pm 1$ corresponds to perfectly circular motion with opposite handedness. The color scale used in Fig.~\ref{fig:phonon_bands_Lz_uniform_masses} represents this quantity, providing a visualization of the chirality of each phonon branch. Finally in Fig.~\ref{fig:chiral_phonon_modes_56} we show the phonon eigenmodes for the chiral bands.
	 
	\subsection{Berry curvature, chiral polarization, and quantized angular momentum}
	\label{subsec:berry_chirality_symmetry}
	
	The existence of the band crossing as shown in Fig.~\ref{fig:phonon_bands_Lz_uniform_masses} near the chiral bands exists for wide range of parameters and so in this section we take a closer look at those. To characterize the topology of the phonon bands, we compute Berry fluxes and Chern numbers on a discretized BZ using the gauge-invariant Fukui-Hatsugai-Suzuki (FHS) construction \cite{fukui2005chern}(see Supplemental Material \cite{kiran2026supplementary} for details).
	
\begin{figure}
		\centering
		\includegraphics[width=\linewidth]{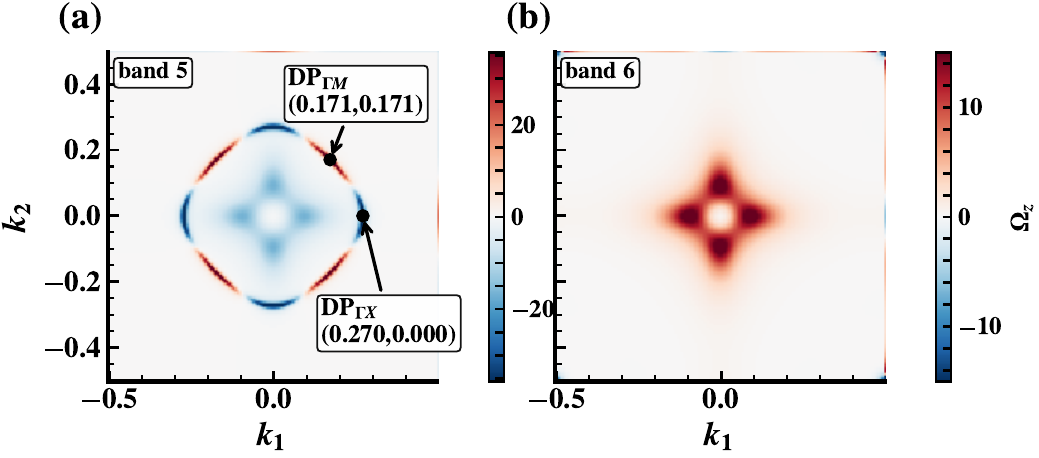}
		\caption{Berry curvature $\Omega_z(\mathbf{k})$ for (a) band 5 and (b) band 6 in the presence of a finite TRS breaking term ($M_z=1.0$) for equal masses $(m_\alpha= 1$) and spring constants $K_1=1.0$, $K_2=0.5$, and $K_3=0.1$. For band 5, the curvature is distributed along a ring-like locus around $\Gamma$ and is enhanced near the Dirac points on the $\Gamma$--$M$ and $\Gamma$--$X$ lines, marked as $\mathrm{DP}_{\Gamma M}$ and $\mathrm{DP}_{\Gamma X}$.}
		\label{fig:berry_curvature_56}
	\end{figure}

	In Fig.~\ref{fig:berry_curvature_56} we show the Berry curvature distributions $\Omega_z(\mathbf{k})$ for the two phonon branches of interest. For band~5, $\Omega_z(\mathbf{k})$ is strongly momentum selective and the curvature is concentrated on an approximately ring-like locus surrounding $\Gamma$, with alternating positive (red) and negative (blue) segments. The dominant weight appears near the crossing points along the $\Gamma-M$ and $\Gamma-X$ directions ($\text{DP}_{\Gamma M}$ and $\text{DP}_{\Gamma X}$), indicating that the geometric response of this branch is controlled primarily by the local band structure in the vicinity of these crossings, where interband hybridization is enhanced and the TRS breaking perturbation produces an avoided crossing. In contrast, band~6 exhibits a more localized, sign-definite curvature texture centered around $\Gamma$, forming a compact fourfold-symmetric pattern consistent with the lattice symmetry and suggesting that the geometric phase of this branch is accumulated predominantly from small-$|\mathbf{k}|$ states. Integrating the curvature over the BZ yields quantized Chern numbers, \begin{equation} C_n=\frac{1}{2\pi}\int_{\mathrm{BZ}} d^2k\,\Omega^{(n)}_z(\mathbf{k}) \end{equation} from which we obtain $C_5\simeq -1$ and $C_6\simeq +1$ for bands~5 and~6, respectively.
	
	To quantify the connection between Berry curvature and chiral phonon motion, we use the circular polarization $S_{z,n}(\mathbf{k})$ defined in Eq.~\ref{eq:sz_phonon} and introduce the Berry-curvature-weighted chiral purity
	\begin{equation}
		P_n^{\Omega}
		=
		\frac{
			\sum_{\mathbf{k}}
			\left|\Omega_{z,n}(\mathbf{k})\right|
			\left|S_{z,n}(\mathbf{k})\right|
		}{
			\sum_{\mathbf{k}}
			\left|\Omega_{z,n}(\mathbf{k})\right|
		}.
		\label{eq:berry_weighted_chiral_purity}
	\end{equation}
	
	This quantity is not a topological invariant and is not expected to be quantized. Instead, it measures how circularly polarized the phonon states are in those regions of the BZ that carry the dominant Berry-curvature weight. This distinction is important because $S_z$ measures the local handedness of the displacement pattern, whereas $\Omega_z$ measures the momentum-space geometry of the eigenvectors.
	
\begin{figure}
		\centering
		\includegraphics[width=\linewidth]{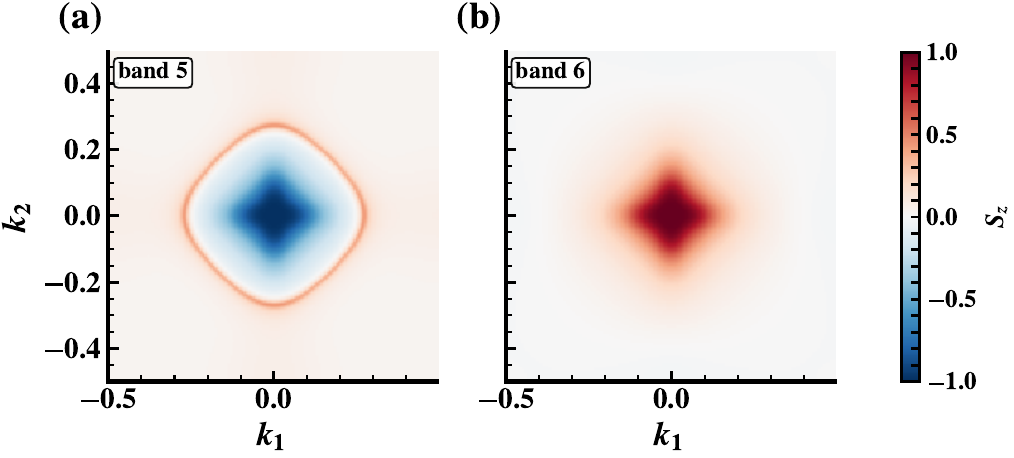}
		\caption{Circular polarization $S_{z,n}(\mathbf{k})$ for (a) band 5 and (b) band 6 in the presence of a finite TRS breaking term ($M_z=1.0$) for equal masses $(m_\alpha= 1$) and spring constants $K_1=1.0$, $K_2=0.5$, and $K_3=0.1$. The regions with enhanced circular polarization occur in the same momentum regions where the Berry curvature is concentrated, although the sign structure of $S_z$ need not follow that of $\Omega_z$.}
		\label{fig:sz_maps_56}
	\end{figure}

	The comparison between Figs.~\ref{fig:berry_curvature_56} and \ref{fig:sz_maps_56} shows that the Berry-curvature hotspots are associated with enhanced circular polarization. For band 5, the Berry curvature changes sign around the ring-like structure, while $S_z$ remains predominantly of one handedness. Thus, the two quantities are correlated in magnitude but are not identical pointwise. The values in Table~\ref{tab:berry_weighted_chiral_purity} make this relation quantitative. Although the BZ average of $|S_{z,n}|$ is small, the Berry-curvature-weighted chiral purity is significantly larger, showing that the states responsible for the Chern number are much more circularly polarized than a typical state in the same band.
	
	It should be noted that $S_{z,n}(\mathbf{k})$ and $\Omega_{z,n}(\mathbf{k})$ characterize different properties of the phonon mode. The circular polarization $S_{z,n}(\mathbf{k})$ measures the real-space handedness of the atomic displacement pattern, whereas the Berry curvature $\Omega_{z,n}(\mathbf{k})$ measures the momentum-space geometry of the eigenvector. Therefore, the sign of $\Omega_{z,n}(\mathbf{k})$ is not fixed solely by the sign of $S_{z,n}(\mathbf{k})$.

	\begin{table}[!ht]
		\centering
		\caption{Berry-curvature-weighted chiral purity for the two topological phonon bands shown in Figs.~\ref{fig:berry_curvature_56} and \ref{fig:sz_maps_56}. The ratio $P_n^{\Omega}/\langle |S_{z,n}| \rangle_{\mathrm{BZ}}$ compares the chirality of the Berry-curvature-carrying states with the average chirality of the full band.}
		\label{tab:berry_weighted_chiral_purity}
		\begin{ruledtabular}
			\begin{tabular}{ccccc}
				Band $n$ & $C_n$ & $\langle |S_{z,n}| \rangle_{\mathrm{BZ}}$ & $P_n^{\Omega}$ & $P_n^{\Omega}/\langle |S_{z,n}| \rangle_{\mathrm{BZ}}$ \tabularnewline
				5 & $-1$ & 0.065 & 0.357 & 5.5 \tabularnewline
				6 & $+1$ & 0.051 & 0.465 & 9.1 \tabularnewline
			\end{tabular}
		\end{ruledtabular}
	\end{table}
	
	We finally relate the chiral character of these modes to quantized angular momentum. At the $C_4$ invariant momenta $\Gamma$ and $M$, the phonon modes can be labeled by their $C_4$ rotation eigenvalues, while at $X$ the relevant little-group eigenvalue is that of $C_2$. We denote these eigenvalues by $\eta_\Gamma$, $\eta_M$, and $\eta_X$. The pseudoangular momentum is defined by
	\begin{equation}
		\eta
		=
		\exp \, \left(-i\frac{2\pi \ell}{4}\right),
		\label{eq:pam_rotation_eigenvalue}
	\end{equation}
	with $\ell=0,\pm1,2$ for a $C_4$ eigenstate.
	
	The results are summarized in Table~\ref{tab:rotation_labels_chern}. Bands 5 and 6 carry opposite $C_4$ eigenvalues at $\Gamma$. Band 5 has $\eta_\Gamma=-i$, corresponding to $\ell_\Gamma=+1$, while band 6 has $\eta_\Gamma=+i$, corresponding to $\ell_\Gamma=-1$. These symmetry labels are accompanied by opposite circular polarization, $S_z(\Gamma)=-1$ and $S_z(\Gamma)=+1$, respectively. Thus, the two topological branches form a counter-rotating pair of chiral optical modes at $\Gamma$.
	
	\begin{table}[!ht]
		\centering
		\caption{Rotation eigenvalues and pseudoangular-momentum labels for the two topological phonon bands. The two bands carry opposite $C_4$ pseudoangular momentum at $\Gamma$, accompanied by opposite circular polarization.}
		\label{tab:rotation_labels_chern}
		\begin{ruledtabular}
			\begin{tabular}{ccccccc}
				Band $n$ &
				$C_n$ &
				$\eta_\Gamma$ &
				$\ell_\Gamma$ &
				$S_z(\Gamma)$ &
				$\eta_M$ &
				$\eta_X$
				\tabularnewline
				5 & $-1$ & $-i$ & $+1$ & $-1$ & $-1$ & $+1$
				\tabularnewline
				6 & $+1$ & $+i$ & $-1$ & $+1$ & $+1$ & $-1$
				\tabularnewline
			\end{tabular}
		\end{ruledtabular}
	\end{table}
	
	The additional eigenvalues at $M$ and $X$ provide a symmetry-based check on the Chern numbers. With the convention in Eq.~\ref{eq:pam_rotation_eigenvalue}, the rotation eigenvalues satisfy
	\begin{equation}
		\exp \, \left(-i\frac{2\pi C_n}{4}\right)
		=
		\eta_\Gamma \eta_M \eta_X .
		\label{eq:chern_rotation_indicator}
	\end{equation}
	For band 5, $\eta_\Gamma \eta_M \eta_X=(-i)(-1)(+1)=+i$, which gives $C_5\equiv 3 \pmod 4$, consistent with $C_5=-1$. For band 6, $\eta_\Gamma \eta_M \eta_X=(+i)(+1)(-1)=-i$, which gives $C_6\equiv 1 \pmod 4$, consistent with $C_6=+1$. 
	
	To further support the bulk Chern-number calculation, we examine the phonon spectrum in a clean strip geometry. The strip is periodic along one lattice direction and open along the other, so that the momentum $k_{\parallel}$ remains a good quantum number. For each value of $k_{\parallel}$, we solve the corresponding strip phonon quadratic eigenvalue problem and analyze the spatial distribution of the eigenmodes across the open direction. The boundary weights are defined as
	\begin{equation}
		W_{\mathrm{top}}
		=
		\frac{
			\sum_{i\in \mathcal{E}_{\mathrm{top}}} |u_i|^2
		}{
			\sum_i |u_i|^2
		},
		\qquad
		W_{\mathrm{bottom}}
		=
		\frac{
			\sum_{i\in \mathcal{E}_{\mathrm{bottom}}} |u_i|^2
		}{
			\sum_i |u_i|^2
		},
	\end{equation}
	where $u_i$ is the physical displacement amplitude of site $i$, and $\mathcal{E}_{\mathrm{top}}$ and $\mathcal{E}_{\mathrm{bottom}}$ denote the boundary rows at the two open edges of the strip. We also define the total edge weight
	\begin{equation}
		W_{\mathrm{edge}}
		=
		W_{\mathrm{top}}+W_{\mathrm{bottom}},
		\label{eq:strip_edge_weight}
	\end{equation}
	which distinguishes boundary-localized modes from bulk-like strip modes. To identify which boundary hosts a given mode, we further introduce the side-resolved edge polarization
	\begin{equation}
		P_{\mathrm{edge}}
		=
		\frac{
			W_{\mathrm{top}}-W_{\mathrm{bottom}}
		}{
			W_{\mathrm{edge}}
		}.
		\label{eq:strip_edge_polarization}
	\end{equation}
	Thus, $P_{\mathrm{edge}}\simeq +1$ corresponds to a mode localized predominantly on the top boundary, while $P_{\mathrm{edge}}\simeq -1$ corresponds to localization on the bottom boundary.
	
	\begin{figure}
		\centering
		\includegraphics[width=\linewidth]{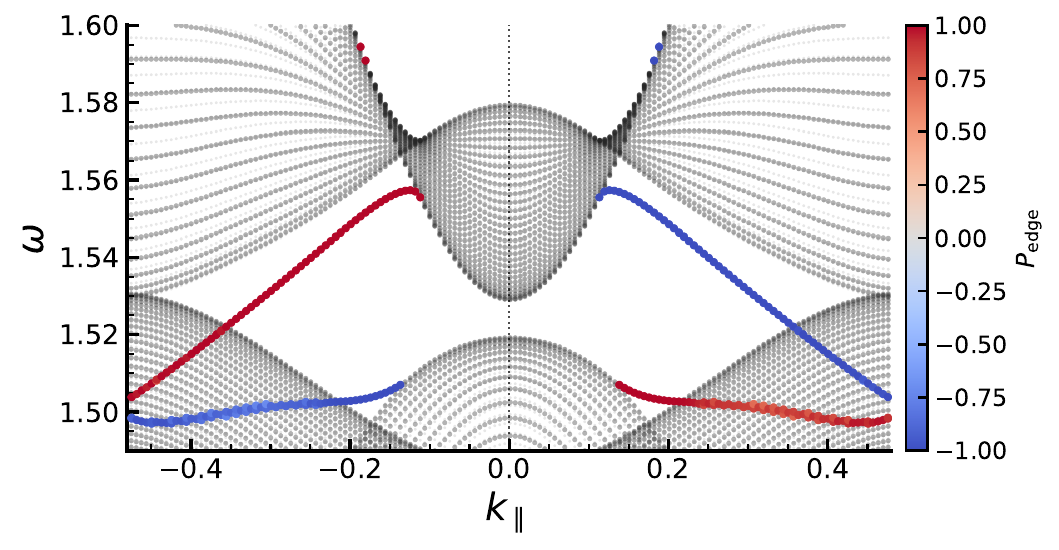}
		\caption[Clean strip spectrum in the projected gap region]
		{Clean strip spectrum in the projected gap region. The strip is periodic along one direction and open along the other. Grey points show the full strip spectrum in the displayed frequency window. Modes with appreciable boundary weight are highlighted and colored by the side-resolved edge polarization $P_{\mathrm{edge}}=(W_{\mathrm{top}}-W_{\mathrm{bottom}})/W_{\mathrm{edge}}$. Red and blue indicate modes localized predominantly on opposite open boundaries. The edge-localized branches lie in the projected gap region and merge into the projected bulk continuum near the edges of the displayed momentum window.}
		\label{fig:strip_edge_spectrum}
	\end{figure}
	
	The resulting strip spectrum is shown in Fig.~\ref{fig:strip_edge_spectrum}. For visual clarity, this calculation is performed at a slightly larger TRS-breaking strength, $M_z=4.0$, which opens a clearer projected gap. We have checked that the relevant bulk Chern numbers remain unchanged in this parameter range. Grey points show the full strip spectrum in the displayed frequency window, while modes with appreciable boundary weight are highlighted and colored according to $P_{\mathrm{edge}}$. The colored branches inside the projected gap are strongly localized near the open boundaries, whereas the dense grey background corresponds mainly to bulk-like strip modes.
	
	The sign of $P_{\mathrm{edge}}$ further shows that the in-gap branches are localized on opposite edges of the strip. In particular, the red and blue branches correspond to modes residing predominantly on the two different boundaries. These edge-localized branches merge into the projected bulk continuum near the edges of the displayed momentum window, as expected for boundary modes of a two-dimensional bulk spectrum projected onto the conserved strip momentum.

	\subsection{Little-group analysis of line crossings}
	\label{subsec:little_group}
	
	A Dirac point on a high-symmetry line arises due to two accidental bands crossings. In that case, each crossing branch must be doubly degenerate and belong to different corepresentations. Recent studies \cite{gong2022dirac, zhang2023encyclopedia} have shown that little-group (co)representations provide a systematic route to identify and classify band degeneracies in low-dimensional crystals. In this section we present the symmetry analysis for existence of the Dirac points along the high-symmetry line $\Gamma-M$ and $\Gamma-X$.	
	
	In the TRS preserved case, ($M_z=0$), the symmetry constraints relevant to the phonon crossings are formulated for the mass-normalized dynamical matrix
	$D_0(\mathbf k)$. For a space-group element $g=(R_g|\tau_g)$ with point operation $R_g$, symmetry implies
	\begin{equation}
		\label{eq:sym_cov_D0}
		U_g(\mathbf k)\,D_0(\mathbf k)\,U_g(\mathbf k)^\dagger
		=
		D_0(R_g\mathbf k)
	\end{equation}
	where $U_g(\mathbf k)$ acts on the $N$-component periodic displacement vector. Along the high-symmetry lines considered here, the relevant operations are symmorphic and act within the home cell, so $U_g(\mathbf k)$ may be taken $\mathbf k$-independent.
	
	Along $\Gamma-X$ ($k_y=0$), the little co-group contains the mirror
	$m_y:(x,y)\mapsto(x,-y)$, represented by
	\begin{equation*}
		\label{eq:U_my}
		U_{m_y}
		=
		P_{m_y}\otimes
		\mathrm{diag}(1,-1)
	\end{equation*}
	where $P_{m_y}$ exchanges $A\leftrightarrow C$ and leaves $B,D$ fixed.
	Along $\Gamma$--$M$ ($k_x=k_y$), the little co-group contains the diagonal mirror
	$m_d:(x,y)\mapsto(y,x)$, represented by
	\begin{equation*}
		\label{eq:U_md}
		U_{m_d}
		=
		P_{m_d}\otimes
		\begin{pmatrix}
			0 & 1\\
			1 & 0
		\end{pmatrix}
	\end{equation*}
	where $P_{m_d}$ exchanges $A\leftrightarrow B$ and $C\leftrightarrow D$.
	
	On a mirror-invariant line one has $ R_g\mathbf k=\mathbf k+\mathbf G$
	with $\mathbf G$ a reciprocal lattice vector. In the present symmorphic setting this reduces to $R_g\mathbf k=\mathbf k$, and therefore $[U_g,D_0(\mathbf k)]=0$. Hence nondegenerate phonon modes on the line can be chosen as simultaneous eigenstates of $D_0(\mathbf k)$ and the mirror operator,
	\begin{equation*}
		\label{eq:mirror_char}
		U_g\,\mathbf v_n(\mathbf k)
		=
		\chi_n(\mathbf k)\,\mathbf v_n(\mathbf k)
		\qquad
		\chi_n=\pm 1
	\end{equation*}
	
	\begin{figure}
		\centering
		\includegraphics[width=\linewidth]{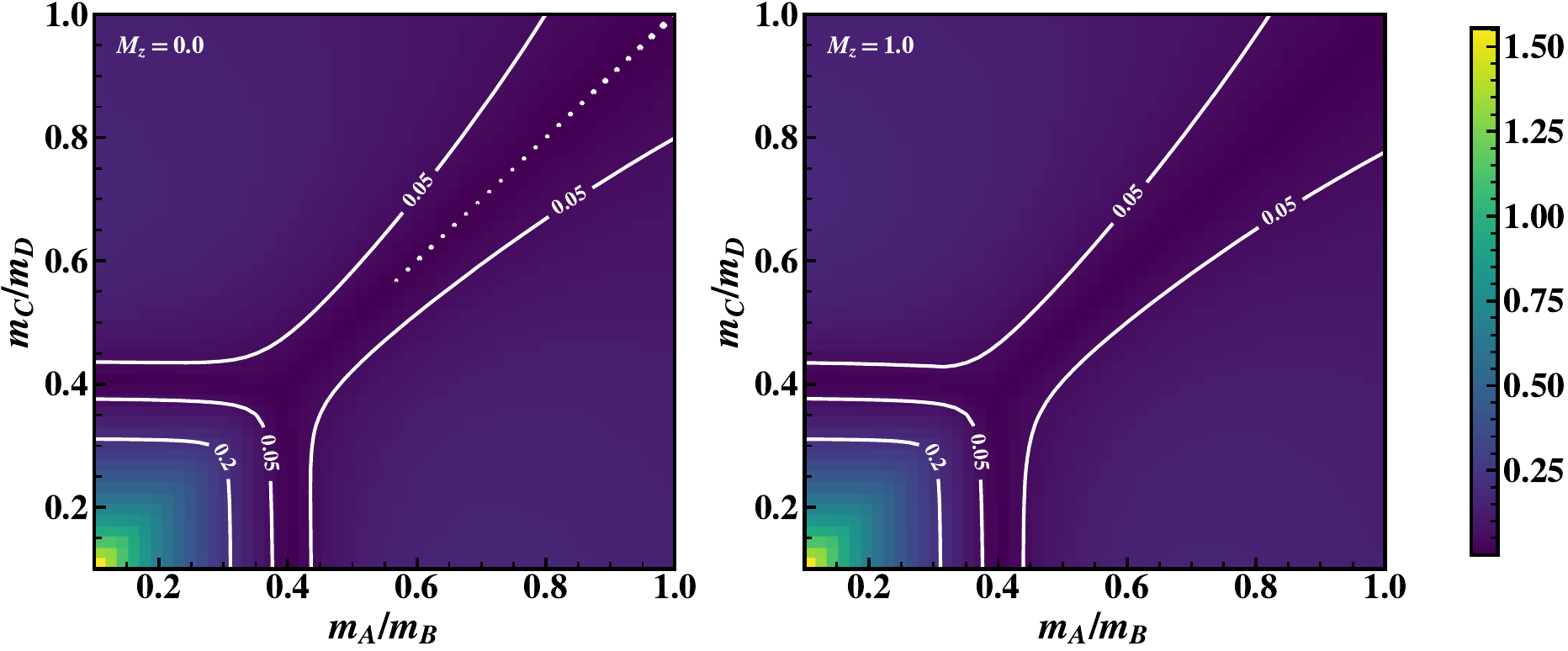}
		\caption{Minimum gap between the two branches forming the local Dirac crossing along $\Gamma$--$X$, plotted in the mass-ratio plane $(m_A/m_B,m_C/m_D)$ for  (a) $M_z=0$ and (b) $M_z=1.0$, with $K_1=1.0$, $K_2=0.5$, and $K_3=0.1$. The white contours indicate representative gap values, while the dot marks points to zero gap (value falls below the numerical threshold).}
		\label{fig:mass_ratio_gap_phase_diagram}
	\end{figure}
	
	The consequence is that any mirror-even operator $O(\mathbf k)$ satisfying $[O(\mathbf k),U_g]=0$ has vanishing matrix elements between opposite mirror sectors i.e, $\mathbf v_n(\mathbf k)^\dagger O(\mathbf k)\,\mathbf v_m(\mathbf k)=0$. Therefore, on the invariant line, two bands with opposite mirror character cannot hybridize through mirror-preserving perturbations. Crossings between opposite mirror sectors are thus symmetry-allowed on the line and, if present, cannot be removed by mirror-preserving hybridization. Their actual occurrence, however, depends on the detailed band ordering and is not enforced by mirror symmetry alone.
	
	The mirror classification further requires that the microscopic parameters preserve the corresponding symmetry. Since the conservative phonon problem is formulated in terms of $D_0(\mathbf k)=M^{-1/2}K(\mathbf k)M^{-1/2}$ the mass matrix must be invariant under the same basis-site permutation. For example, a mass pattern $m_A=m_C\neq m_B=m_D$ preserves $m_y$ but generally breaks $m_d$. In that case, mirror eigenvalues remain well defined along $\Gamma$--$X$, but the diagonal-mirror classification on $\Gamma$--$M$ is lost and an otherwise allowed crossing there can become an avoided crossing.
	
	Incorporating the mirror odd and TRS breaking term through the Hermitian kernel $i\widetilde A(\mathbf k)$ into the quadratic phonon problem provides the leading symmetry-allowed mixing between the $\chi=\pm1$ sectors and generically gaps the crossing.

	To make the symmetry argument explicit, in Fig.~\ref{fig:mass_ratio_gap_phase_diagram} we map the minimum $\Gamma$--$X$ gap in the mass-ratio plane. In the $M_z=0$ case, the nearly gapless locus lies close to the diagonal condition $m_A/m_B \simeq m_C/m_D$, which is the natural constraint for preserving the $m_y$ mirror. Along this line the two crossing branches retain well-defined and opposite mirror characters, so the crossing remains symmetry-allowed and cannot be removed by mirror-preserving hybridization. Once the mirror-odd TRS-breaking term is turned on, the same region of parameter space no longer supports a zero gap.
	
	\subsection{Effect of TRS breaking on local mode character}
	\label{subsec:chirality_transfer}
	
	\begin{figure}
		\centering
		\includegraphics[width=\linewidth]{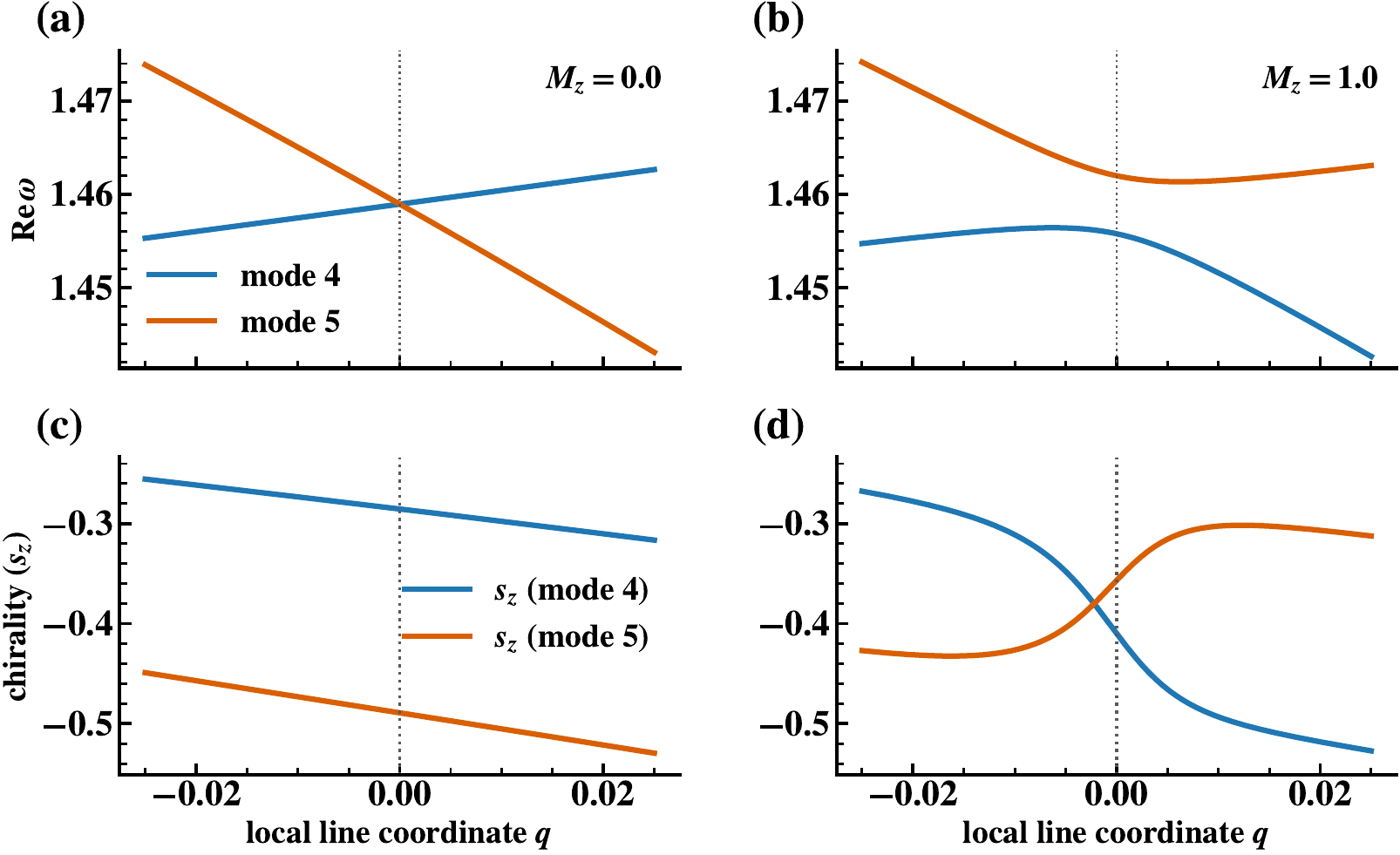}
		\caption{Local mode evolution along the $\Gamma$--$X$ line for the two modes associated with the band crossing. (a, b) $\mathrm{Re}\,\omega$ and (c, d) the corresponding mass-weighted chirality $s_z$, with the two branches tracked by wavefunction overlap so that the mode character is followed continuously through the crossing region. For $M_z=0$, the two tracked modes cross without appreciable change in their chirality, consistent with negligible mixing of the underlying mode characters. For finite $M_z$, the crossing is replaced by an avoided crossing, accompanied by a smooth transfer of chirality between the two tracked branches.}
		\label{fig:chirality_transfer_GX}
	\end{figure}
	
	To further elucidate the band-4/band-5 crossings, we analyze the local evolution of the two relevant modes along the $\Gamma$--$X$ and $\Gamma$--$M$ directions, tracking the branches continuously by wavefunction overlap through the crossing region. The local coordinate $q$ is measured relative to the point of minimum gap on the corresponding high-symmetry line, so that $q=0$ coincides with the crossing point for $M_z=0$ and with the center of the avoided crossing for finite $M_z$.
	
	\begin{figure}
		\centering
		\includegraphics[width=\linewidth]{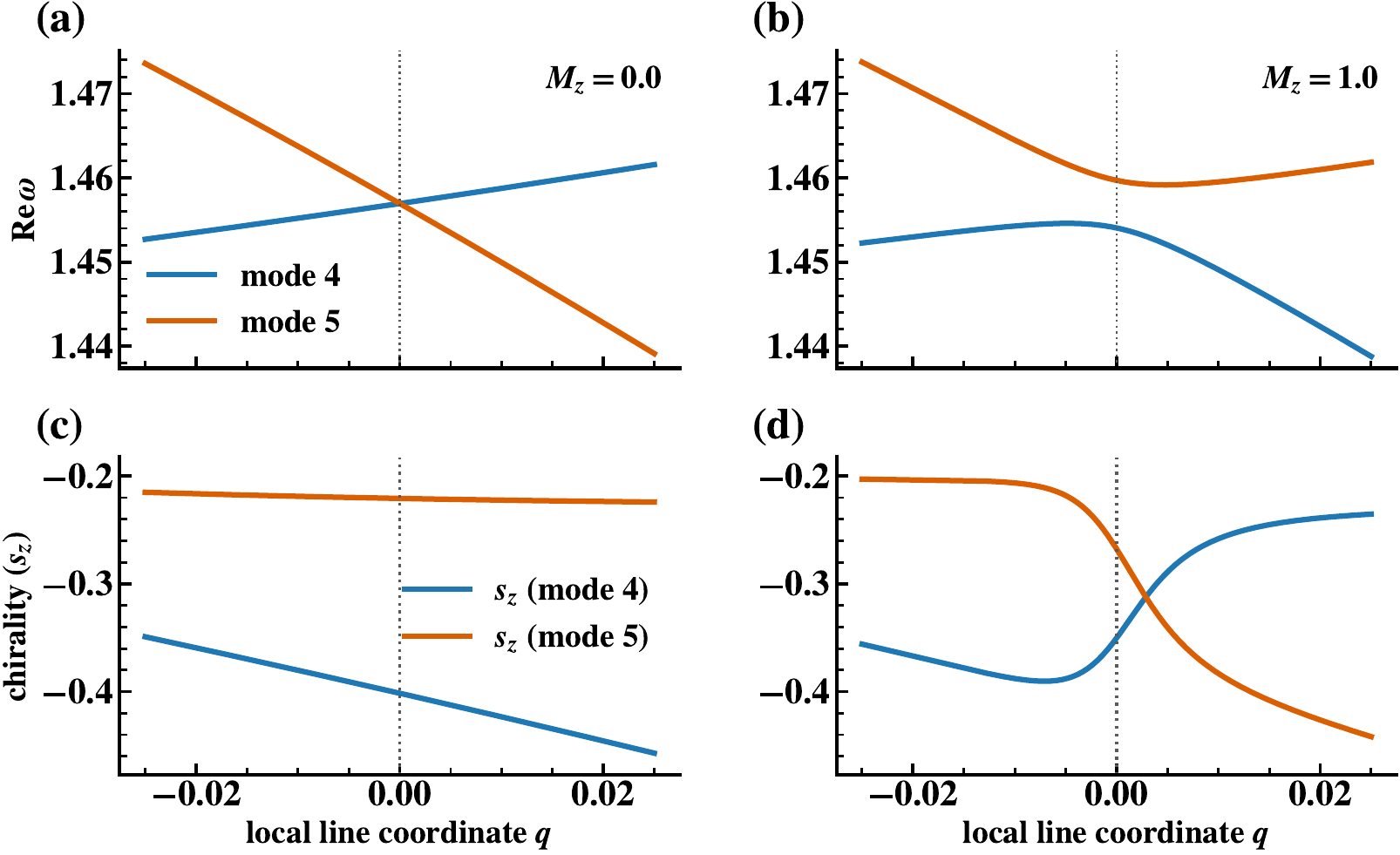}
		\caption{Local mode evolution along the $\Gamma$--$M$ line for the two modes associated with the band crossing. (a,b) $\mathrm{Re},\omega$ and (c,d) the corresponding circular polarization $S_z$, with the two branches tracked by wavefunction overlap so that the mode character is followed continuously through the crossing region. For $M_z=0$, the two tracked modes cross without appreciable change in their circular polarization, consistent with negligible mixing of the underlying mode characters. For finite $M_z$, the crossing is replaced by an avoided crossing, accompanied by a smooth transfer of $S_z$ between the two tracked branches.}
		\label{fig:chirality_transfer_GM}
	\end{figure}
	
	As shown in Figs.~\ref{fig:chirality_transfer_GX} and \ref{fig:chirality_transfer_GM}, for $M_z=0$ the two tracked modes cross without any appreciable change in their circular polarization $S_z$ along either direction. This indicates that each branch retains its mode character through the degeneracy and that the hybridization between the two modes remains negligible. For finite $M_z$, however, the crossing is replaced by a clear avoided crossing in both directions, accompanied by a smooth exchange of $S_z$ between the two overlap-tracked branches. The TRS-breaking perturbation lifts the degeneracy and mixes the two otherwise distinct modes, thereby redistributing their circular-polarization character across the avoided-crossing region.
	
\begin{figure}
		\centering
		\includegraphics[width=\linewidth]{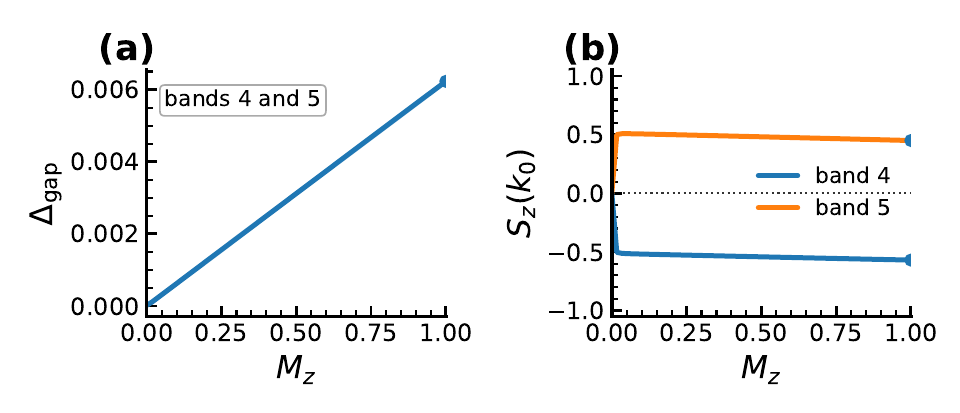}
		\caption{Evolution of the TRS-breaking gap and local chiral character near the avoided crossing on the $\Gamma$--$X$ line.
			(a) Direct gap $\Delta_{\mathrm{gap}}$ between bands 4 and 5 at the crossing point $k_0$ as a function of the TRS-breaking parameter $M_z$.
			(b) Circular polarization $S_z(k_0)$ of the two hybridizing branches.
			For finite $M_z$, the gap opens and the two modes acquire opposite circular polarization, showing that the local gap-opening process is accompanied by chiral hybridization of the phonon modes.}
		\label{fig:gap_chirality_evolution}
	\end{figure}

	To make this gap-opening process more explicit, we also track the direct gap and the circular polarization at the crossing point $k_0$ as a function of $M_z$. The result is shown in Fig.~\ref{fig:gap_chirality_evolution}. The direct gap $\Delta_{\mathrm{gap}}$ opens continuously from zero and grows approximately linearly with $M_z$, consistent with the TRS-breaking term acting as the local gap-opening perturbation. At the same time, the two hybridized branches acquire opposite values of $S_z(k_0)$, showing that the opening of the gap is accompanied by the formation of counter-rotating local phonon character.
	
	We emphasize that the extremal points of the hybridized branches need not occur exactly at $q=0$, even though the minimum gap is centered there. This offset reflects the fact that the avoided crossing is superimposed on a locally tilted average dispersion, so that the upper and lower adiabatic branches develop their turning points on opposite sides of the minimum-gap point. Although the circular polarization does not necessarily reverse sign across the avoided crossing, the smooth interchange of $S_z$ and the finite-$M_z$ splitting shown in Fig.~\ref{fig:gap_chirality_evolution} together provide a direct local signature of the TRS-induced mixing mechanism. This establishes the local sequence by which TRS breaking converts the crossing into a chiral avoided crossing, which then contributes to the Berry-curvature hotspots and the topological character discussed in Sec.~\ref{subsec:berry_chirality_symmetry}.
	
	\subsection{Disorder response of the $\Gamma$-point chiral phonons}

	To study the stability of the chiral phonon pair identified at $\Gamma$ in the clean square-octagon lattice, we construct a finite $L_x\times L_y$ real-space supercell with periodic boundary conditions. Static disorder is introduced directly in real space through random site masses and/or random bond stiffnesses \cite{monthus2010anderson}. For a given supercell realization, we take

	\begin{equation}
		m_i=m_i^{(0)}(1+\eta_i),\qquad \eta_i\in[-W,W],
	\end{equation}
	and
	\begin{equation}
		K_\ell=K_\ell^{(0)}(1+\xi_\ell),\qquad \xi_\ell\in[-W,W],
	\end{equation}
	
	$W$ is a dimensionless relative disorder amplitude. The random variables entering the site masses and bond stiffnesses are independently drawn from uniform distributions, $\eta_i,\xi_\ell\in[-W,W]$, where $\eta_i$ represents the relative fluctuation of the mass at site $i$, and $\xi_\ell$ represents the relative fluctuation of the spring constant on bond $\ell$. Here $i,j$ label sites in the disordered supercell and $\ell$ labels bonds. Mass disorder models local inertial mismatch, while spring disorder models spatial fluctuations of the interatomic force constants.
	
	For each realization, we construct the real-space supercell matrices $D_0^{\mathrm{SC}}$ and $iA^{\mathrm{SC}}$ and solve
	\begin{equation}
		\big[D_0^{\mathrm{SC}}-\omega^2 I+\omega\, iA^{\mathrm{SC}}\big]v=0 .
	\end{equation}
	To follow the clean chiral sector under disorder, we use the two clean $\Gamma$-point eigenvectors as a reference subspace. After tiling these vectors uniformly across the supercell and orthonormalizing them, we obtain a two-dimensional basis $Q_{\mathrm{ref}}$ and the corresponding projector, $P_{\mathrm{ref}}=Q_{\mathrm{ref}}Q_{\mathrm{ref}}^\dagger$. For each normalized supercell eigenmode $U_n$, the overlap weight $w_n=\langle U_n|P_{\mathrm{ref}}|U_n\rangle$ measures how strongly that mode belongs to the clean $\Gamma$-point chiral subspace. We use these weights to define projector-weighted observables in a fixed frequency window $\mathcal W$ centered on the clean pair,
	\begin{equation}
		\overline{X}
		=
		\frac{\sum_{n\in\mathcal W} w_n X_n}{\sum_{n\in\mathcal W} w_n},
	\end{equation}
	
	where $X_n$ is the mode-resolved quantity of interest. This construction remains meaningful even when disorder spreads the clean-pair character over several nearby supercell modes. Thus, the projector tracks the survival of the clean $\Gamma$-point sector as a subspace, while the chirality diagnostics defined below test how much of its circular polarization structure is retained.
	
	The chirality of a supercell mode is evaluated from the physical displacement field $u_i=v_i/\sqrt{m_i}$. Writing $u_i=(u_{ix},u_{iy})$, the global circular polarization is
	\begin{equation}
		\chi_{\mathrm{g}}
		=
		\frac{-2\,\mathrm{Im}\sum_i u_{ix}^*u_{iy}}
		{\sum_i \left(|u_{ix}|^2+|u_{iy}|^2\right)} .
	\end{equation}
	
	The quantity $\chi_{\mathrm{g}}$ is the real-space supercell analogue of the Bloch-mode circular polarization $S_{z,n}(\mathbf{k})$ defined in Eq.~(11). In the clean periodic system, translational symmetry allows the circular polarization to be assigned to each band $n$ at each crystal momentum $\mathbf{k}$. In the disordered supercell, $\mathbf{k}$ is no longer a good quantum number, and the corresponding global circular polarization is evaluated by summing over all sites of the supercell. The notation $\chi_{\mathrm g}$ is used to distinguish this net supercell chirality from the local chirality measure $\chi_{\mathrm{loc}}$ defined below.
	
	\begin{figure}
		\centering
		\includegraphics[width=\linewidth]{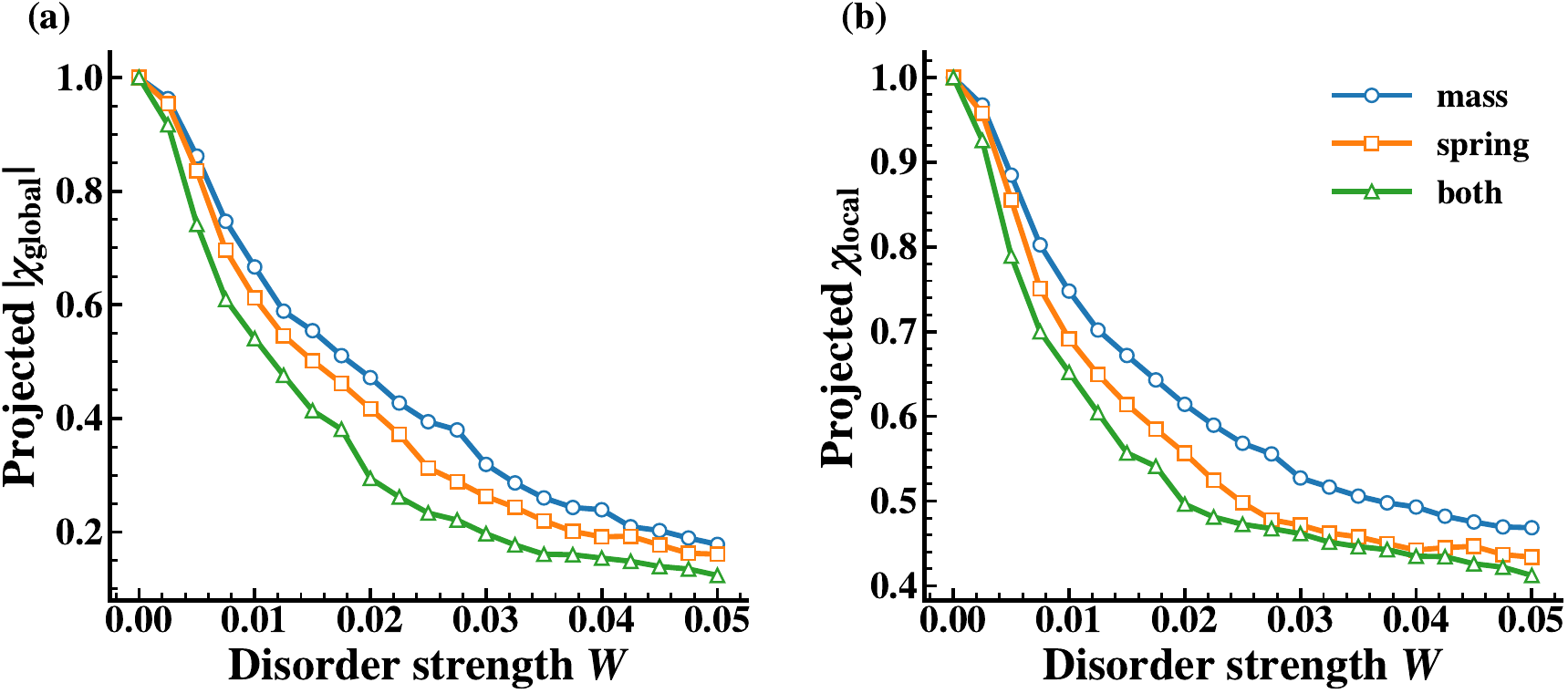}
		\caption{Projector-weighted chirality diagnostics for the $\Gamma$-point chiral phonon sector in the square-octagon lattice. (a) The projected mean global chirality and (b) the projected mean local chirality, for mass disorder, spring-constant disorder, and the combined case. These quantities measure the degradation of the polarization structure within the tracked clean-sector subspace.}
		\label{fig:projected_chirality_global_local}
	\end{figure}
	
	Since the two clean partner modes carry opposite handedness, the relevant projected global measure is the projector-weighted mean of $|\chi_{\mathrm g}|$. To distinguish genuinely nonchiral motion from locally chiral motion that cancels globally, we also define the local chirality density $c_i=-2\,\mathrm{Im}(u_{ix}^*u_{iy})$, and the associated local chirality strength
	\begin{equation}
		\chi_{\mathrm{loc}}
		=
		\frac{\sum_i |c_i|}
		{\sum_i \left(|u_{ix}|^2+|u_{iy}|^2\right)} .
	\end{equation}
	In addition, we monitor the participation ratio,
	\begin{equation}
		\mathrm{PR}=\frac{1}{N_{\mathrm{s}}\sum_i p_i^2},
		\qquad
		p_i=\frac{|u_{ix}|^2+|u_{iy}|^2}{\sum_j (|u_{jx}|^2+|u_{jy}|^2)},
	\end{equation}
	which measures the spatial extent of the mode. To quantify the spectral retention of the clean chiral sector, we also define the window fraction
	\begin{equation}
		F_{\mathrm{win}}
		=
		\frac{\sum_{n\in\mathcal W} w_n}{\sum_n w_n},
	\end{equation}
	which measures how much of the clean-pair weight remains inside the chosen frequency window. A value of $F_{\mathrm{win}}$ close to unity indicates spectral concentration of the clean-sector weight.
	
	Figure~\ref{fig:projected_chirality_global_local} shows that both projected chirality measures decrease monotonically with increasing disorder strength. Mass disorder is least disruptive, spring-constant disorder produces a stronger reduction, and the combined case suppresses the chiral response most rapidly. At the same time, the projected local chirality remains systematically larger than the projected global chirality. Disorder therefore degrades coherent net circular polarization before it eliminates locally chiral motion.
	
	\begin{figure}
		\centering
		\includegraphics[width=\linewidth]{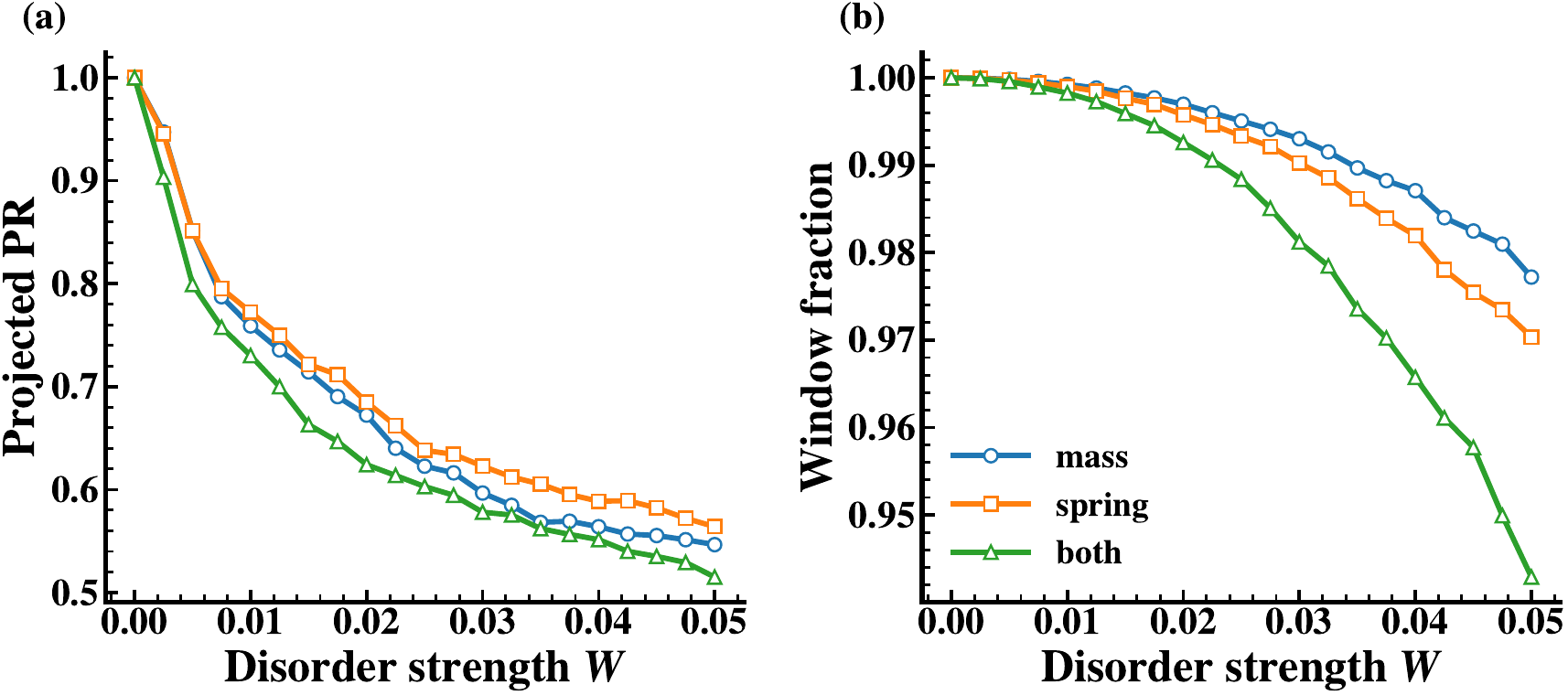}
		\caption{Projector-weighted spatial and spectral diagnostics for the $\Gamma$-point chiral sector. (a) The projected participation ratio and (b) the window fraction $F_{\mathrm{win}}$, measuring the fraction of clean-pair weight retained inside the selected frequency window. The participation ratio decreases with disorder, reflecting increasing spatial inhomogeneity, whereas the window fraction remains close to unity over the full range studied, showing that the clean chiral sector remains spectrally concentrated even as its polarization structure is progressively degraded.}
		\label{fig:projected_PR_window}
	\end{figure}
	
	In Fig.~\ref{fig:projected_PR_window} we present the projected participation ratio which decreases with $W$. The clean chiral sector becomes progressively more spatially inhomogeneous under disorder. By contrast, the window fraction remains close to unity throughout the studied range, although it is reduced most strongly in the combined-disorder case. Disorder does not immediately expel the clean $\Gamma$-point chiral sector from its original spectral window. Rather, the spectral support of the pair remains well defined while its globally coherent chiral polarization is progressively weakened.
	
	\begin{figure}[!ht]
		\centering
		\includegraphics[width=\linewidth]{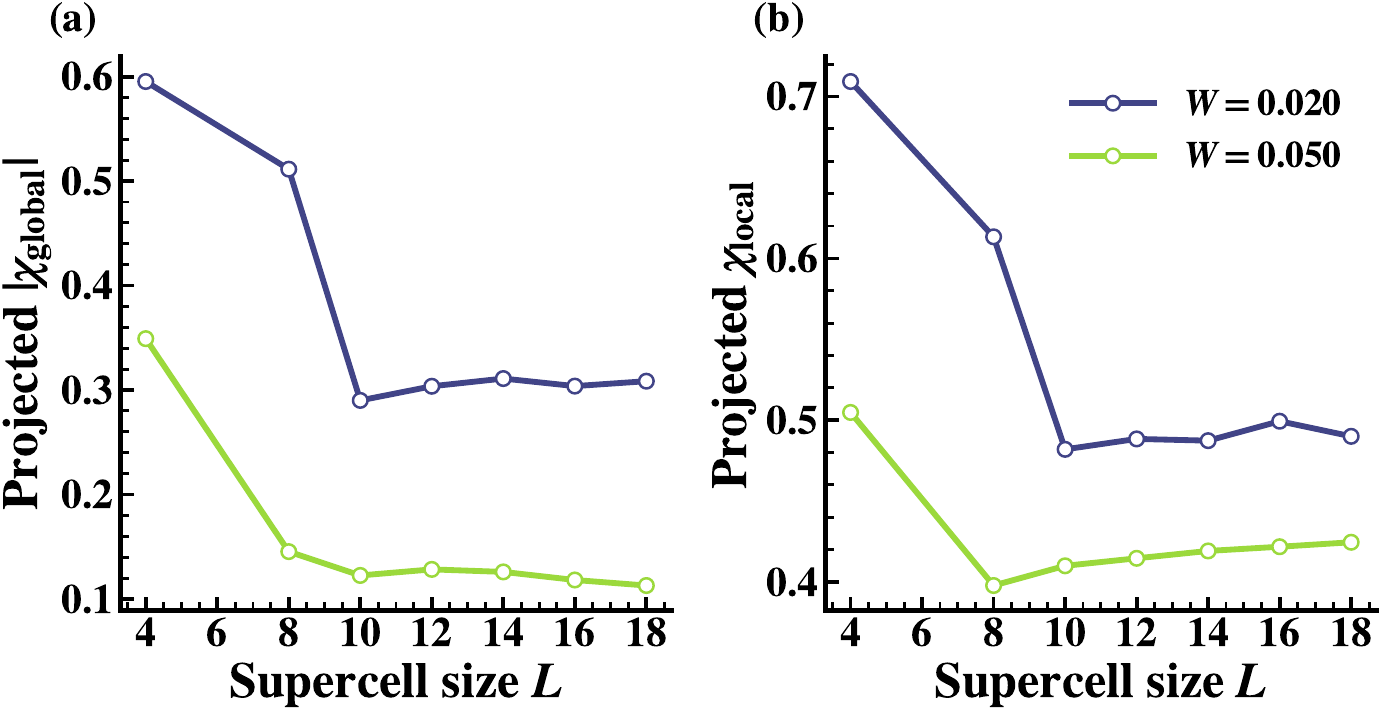}
		\caption{Supercell-size dependence of the projector-weighted chirality diagnostics for the combined-disorder case. (a) The projected mean global chirality and (b) the projected mean local chirality for two representative disorder strengths.}
		\label{fig:projected_chirality_convergence}
	\end{figure}
	
	For the disorder analysis, we fix $M_z=1.0$ and use $K_1=1.0$, $K_2=0.5$, and $K_3=0.1$. The main disorder-dependent quantities are evaluated in an $L_x=L_y=12$ supercell and averaged over $20$ disorder realizations. To verify that the reported trends are not artifacts of this particular choice, we examine the supercell-size dependence of the projected chirality diagnostics in Fig.~\ref{fig:projected_chirality_convergence}. For $L\gtrsim 10$ both the projected global and local chirality diagnostics show only weak residual dependence on $L$.
		
	The relatively small saturated values in Fig.~\ref{fig:projected_chirality_convergence} arise because this figure corresponds to the combined-disorder case, where both masses and spring constants fluctuate. This is the most disruptive disorder case considered, as also seen from the green curves in Fig.~\ref{fig:projected_chirality_global_local}. In addition, the clean $\Gamma$-point chiral sector consists of two opposite-handed partner modes. Disorder can mix these partners within the same projected subspace, so a disordered mode may retain clean-pair phonon character while its net global chirality is reduced by cancellation between opposite-handed components.
		
	Thus, the disorder response is twofold. The clean $\Gamma$-point phonon sector remains spectrally identifiable and retains finite local circular motion under weak disorder, but its globally coherent chiral polarization is progressively degraded.
	
	\subsection{Model scattering intensity and experimental visibility}
	
	To connect the calculated phonon modes with possible momentum-resolved probes, we evaluate a model one-phonon scattering intensity. Similar eigenvector-resolved spectral intensities are routinely used to interpret inelastic X-ray and neutron scattering measurements of phonons \cite{baron2009phonons,fair2022euphonic}. In the present work, this quantity is not intended as a material-specific cross section, but provides a useful estimate of which phonon branches carry finite spectral weight in an inelastic X-ray, neutron, or EELS-type measurement. For a momentum transfer $\mathbf Q=\mathbf G+\mathbf q$, we use
	\begin{equation}
		S(\mathbf Q,\omega)
		=
		\sum_{\nu}
		I_{\nu}(\mathbf Q)
		G_{\sigma}
		\left(
		\omega-\omega_{\nu\mathbf q}
		\right),
		\label{eq:model_scattering_intensity}
	\end{equation}
	where
	\begin{equation}
		G_{\sigma}
		\left(
		\omega-\omega_{\nu\mathbf q}
		\right)
		=
		\frac{1}{\sqrt{2\pi}\sigma}
		\exp
		\left[
		-\frac{
			\left(
			\omega-\omega_{\nu\mathbf q}
			\right)^2
		}{
			2\sigma^2
		}
		\right]
	\end{equation}
	is a Gaussian broadening function, and $\omega_{\nu\mathbf q}$ is the phonon frequency of branch $\nu$. The mode intensity is taken as \cite{dove2003structure,togo2015first,baron2009phonons}
	\begin{equation}
		I_{\nu}(\mathbf Q)
		=
		\frac{1}{\omega_{\nu\mathbf q}}
		\left|
		\sum_{\kappa}
		f_{\kappa}(\mathbf Q)\,
		\mathbf Q\cdot \mathbf u_{\kappa\nu}(\mathbf q)\,
		e^{i\mathbf Q\cdot \boldsymbol{\tau}_{\kappa}}
		\right|^2 .
		\label{eq:model_mode_intensity}
	\end{equation}
	Here $\mathbf u_{\kappa\nu}(\mathbf q)$ is the physical displacement eigenvector of atom $\kappa$ in phonon branch $\nu$, $\boldsymbol{\tau}_{\kappa}$ is the basis position, and $f_{\kappa}(\mathbf Q)$ is an effective scattering weight. In the present model calculation we take $f_{\kappa}=1$ for all basis sites, since our aim is to identify the eigenvector and basis-interference contribution to the spectral visibility rather than to predict absolute intensities for a specific material.
	
\begin{figure}[!ht]
		\centering
		\includegraphics[width=0.72\linewidth]{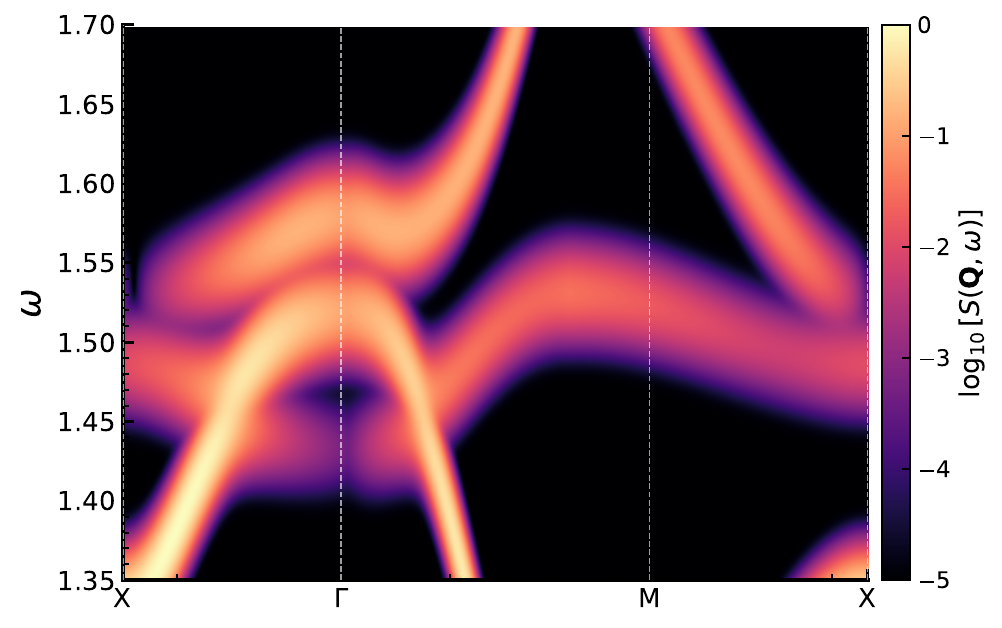}
		\caption[Model one-phonon scattering intensity]
		{Model one-phonon scattering intensity $S(\mathbf Q,\omega)$ in the optical frequency window containing the chiral phonon branches. The intensity is computed from the phonon eigenvectors using Eqs.~\eqref{eq:model_scattering_intensity} and \eqref{eq:model_mode_intensity}. Bright features denote branches with appreciable model scattering weight for the chosen momentum transfer $\mathbf Q=\mathbf G+\mathbf q$.}
		\label{fig:model_scattering_intensity}
	\end{figure}

	The resulting intensity map is shown in Fig.~\ref{fig:model_scattering_intensity}. Bright regions indicate phonon branches with appreciable model scattering weight for the chosen momentum transfer. The optical branches in the frequency range where the chiral and topological features occur retain finite spectral intensity, indicating that these modes are present in the eigenvalue spectrum and should also be accessible to momentum-resolved phonon probes once a material-specific realization is specified.
	
	Chiral phonon modes have been detected using transient infrared spectroscopy in monolayer WSe$_2$ \cite{zhu2018observation}, circularly polarized Raman scattering in $\alpha$-HgS \cite{ishito2023truly}, and circularly polarized resonant inelastic X-ray scattering in quartz \cite{ueda2023chiral}. These examples show that helicity-sensitive optical probes and momentum-resolved scattering techniques provide route for detecting phonon circular polarization. We would like to emphasize that in our model, the absolute intensity and the physical frequency scale would require material-specific atomic form factors, Born effective charges or Raman tensors, and microscopic force constants and masses.
	
	\section{Summary and Conclusion}
	\label{sec:summary_conclusion}

	In this work, we have shown that the square-octagon lattice provides a versatile setting in which lattice geometry, interaction range, mass distribution, and TRS breaking act as distinct control parameters for phonon structure. In particular, the third-nearest-neighbor coupling plays a qualitatively important role in activating optical branches that remain nearly frozen when only shorter-range couplings are included. Physically, this reflects the fact that in a multi-site square-octagon basis, local force pathways alone do not efficiently couple all internal relative motions, whereas longer range bonds promote additional restoring channels that enhance the dispersion and hybridization of optical modes. We further find that mass contrast and force-constant contrast provide an additional route for controlling the vibrational dynamics by redistributing mode character across the lattice and tuning the phononic gaps. These features make the system a promising framework for engineering phonon spectra and band gaps in acoustic and phononic metamaterials.
	
	By introducing a magnetoelastic TRS breaking term, we drive the system into a regime with chiral phonons and analyze the Berry curvature concentrated near the gapped crossings. To understand the symmetry origin of these crossings, we carry out a little-group analysis along both the $\Gamma$--$X$ and $\Gamma$--$M$ directions. In the TRS-preserving case ($M_z=0$), the relevant branches cross with essentially no change in mode character, indicating negligible mixing. Once $M_z$ is finite, these crossings are converted into avoided crossings, with a smooth transfer of mass-weighted chirality between the hybridized branches. We also supplemented the bulk Berry-curvature and Chern-number analysis with a strip-geometry calculation. 
	
	We also examined the robustness of the $\Gamma$-point chiral sector against static disorder introduced through random site masses and random bond stiffnesses. The chiral response is progressively reduced with increasing disorder, with mass disorder producing the weakest suppression, spring disorder a stronger effect, and combined disorder the largest reduction. At the same time, the projected local chirality remains consistently larger than the projected global chirality, while the spectral weight associated with the clean chiral pair stays concentrated within the same frequency window throughout the disorder range studied. These results show that weak to moderate disorder does not immediately erase the chiral sector itself, but instead first degrades its coherent net polarization through increased spatial inhomogeneity and mode mixing. In this sense, the local chiral character is more robust than the globally coherent chiral response.
	
	We also examined the disorder response of the $\Gamma$-point chiral sector by introducing static disorder through random site masses and random bond stiffnesses. The globally coherent chiral response is progressively reduced with increasing disorder, with mass disorder producing the weakest suppression, spring disorder a stronger effect, and combined disorder the largest reduction. At the same time, the projected local chirality remains consistently larger than the projected global chirality, while the spectral weight associated with the clean chiral pair stays concentrated within the same frequency window throughout the disorder range studied. These results show that weak to moderate disorder does not immediately erase the clean $\Gamma$-point chiral sector itself, but instead first degrades its coherent net circular polarization through increased spatial inhomogeneity and mode mixing between the opposite-handed partner modes.
	
	Finally, we evaluated a model one-phonon scattering intensity to estimate the experimental visibility of the relevant modes. Although this quantity is not intended as a material-specific cross section, it shows that the optical branches associated with the chiral and topological features carry finite spectral weight and should therefore be accessible, in principle, to momentum-resolved phonon probes such as inelastic X-ray, neutron, or EELS-type measurements once a concrete material realization is specified.

	\section{Acknowledgments}
	The authors thank the anonymous referees for their careful reading of the manuscript and for their constructive comments, which helped improve the clarity and presentation of the work. R.K. sincerely thanks Subhajit Pramanick for insightful discussions on the topic.
	
\bibliography{references}
\end{document}


\title{Supplementary Material for:\\
		\textit{Chiral and Topological Phonons in the Square-Octagon Lattice}}
	
	\author{Ravi Kiran}
	\email{ravieroy123@iitkgp.ac.in}
	\affiliation{Department of Physics, Indian Institute of Technology Kharagpur, Kharagpur 721302, West Bengal, India}
	
	\author{A. Taraphder}
	\email{arghya@phy.iitkgp.ac.in}
	\affiliation{Department of Physics, Indian Institute of Technology Kharagpur, Kharagpur 721302, West Bengal, India}
	
	\maketitle
	
	\setcounter{tocdepth}{1}
	\tableofcontents
	
	\vspace{1em}
	
	\newpage
	
	\renewcommand{\thesection}{S\arabic{section}}
	\renewcommand{\thesubsection}{S\arabic{section}.\arabic{subsection}}
	\renewcommand{\theequation}{S\arabic{equation}}
	
	\setcounter{section}{0}
	\setcounter{equation}{0}

	\section{Origin of the velocity-dependent kernel}
	
	The velocity-dependent term used in the main text can be motivated from an effective quadratic Lagrangian for the lattice displacements. For compactness, let $\mathbf u$ denote the vector containing all displacement components. We consider
	\begin{equation*}
		L
		=
		\frac{1}{2}\dot{\mathbf u}^{T}M\dot{\mathbf u}
		-
		\frac{1}{2}\mathbf u^{T}K\mathbf u
		-
		\frac{1}{2}\mathbf u^{T}A\dot{\mathbf u},
		\label{eq:S_eff_lagrangian}
	\end{equation*}
	where $M$ is the mass matrix, $K=K^T$ is the harmonic force-constant matrix, and $A$ is a real antisymmetric matrix. The last term is first order in time and is the lattice analogue of a gyroscopic or Lorentz-force coupling. Using the Euler-Lagrange equation and $A^T=-A$ gives,
	
	\begin{equation*}
		\frac{\partial L}{\partial \dot{\mathbf u}}
		=
		M\dot{\mathbf u}
		+
		\frac{1}{2}A\mathbf u,
		\qquad
		\frac{\partial L}{\partial \mathbf u}
		=
		-K\mathbf u
		-
		\frac{1}{2}A\dot{\mathbf u}.
	\end{equation*}
	Therefore,
	\begin{equation*}
		M\ddot{\mathbf u}+A\dot{\mathbf u}+K\mathbf u=0.
		\label{eq:S_eom_from_lagrangian}
	\end{equation*}
	Thus, an antisymmetric first-order-in-time term in the effective Lagrangian directly produces a velocity-dependent contribution to the lattice equations of motion. For a periodic lattice with basis atoms, the same structure becomes
	
	\begin{equation}
		\sum_{R',j}
		\left[
		M_{ij}\delta_{RR'}\partial_t^2
		+
		\mathcal A_{ij}(R-R')\partial_t
		+
		\mathcal K_{ij}(R-R')
		\right]
		\mathbf u_{R',j}(t)=0.
		\label{eq:S_real_space_A_origin}
	\end{equation}
	The nondissipative character of the velocity kernel requires $ \mathcal A_{ij}(R) = -\mathcal A_{ji}(-R)^T$, which ensures that the Bloch-space matrix satisfies $A(\mathbf k)^\dagger=-A(\mathbf k)$.
	
	\section{Low-energy motivation for the TRS-breaking scale \(g\)}
	
		To motivate the phenomenological choice of the time-reversal-symmetry breaking scale $g$, we consider a low-energy description in which the phonon sector is coupled to an auxiliary gapped magnetic fluctuation. The purpose of this construction is not to provide a fully microscopic spin-lattice derivation, but rather to show why a term of the form $g \sim \lambda^2 M_z/\Omega_m$ is a natural scaling form for the effective phonon kernel used in the main text.
		
		We denote by $v(\mathbf{k},\omega)$ the mass-normalized phonon coordinate and by $\phi(\mathbf{k},\omega)$ a collective magnetic fluctuation about a background with out-of-plane time-reversal-odd polarization $M_z$. At the quadratic level, a schematic coupled theory may be written in frequency space as
		\begin{equation}
			\begin{aligned}
				\mathcal{S}
				= {}&
				\frac{1}{2}
				\sum_{\mathbf{k},\omega}
				v^\dagger(\mathbf{k},\omega)
				\big[D_0(\mathbf{k})-\omega^2 I\big]
				v(\mathbf{k},\omega) +
				\frac{1}{2}
				\sum_{\mathbf{k},\omega}
				\phi^\dagger(\mathbf{k},\omega)\,
				\chi_m^{-1}(\mathbf{k},\omega)\,
				\phi(\mathbf{k},\omega)
				\\[4pt]
				&+
				\lambda
				\sum_{\mathbf{k},\omega}
				\Big[
				v^\dagger(\mathbf{k},\omega)\,B(\mathbf{k})\,\phi(\mathbf{k},\omega)
				+
				\phi^\dagger(\mathbf{k},\omega)\,B^\dagger(\mathbf{k})\,v(\mathbf{k},\omega)
				\Big].
			\end{aligned}
		\end{equation}
		Here $B(\mathbf{k})$ encodes the bond and geometry-dependent structure of the magnetoelastic coupling, $\lambda$ is its overall strength, and $\chi_m(\mathbf{k},\omega)$ is the dynamical susceptibility of the gapped magnetic mode.
		
		Integrating out the magnetic fluctuation $\phi$ generates an effective phonon self-energy
		\begin{equation}
			\Sigma_{\mathrm{ph}}(\mathbf{k},\omega)
			=
			-\lambda^2
			B(\mathbf{k})\,
			\chi_m(\mathbf{k},\omega)\,
			B^\dagger(\mathbf{k}).
		\end{equation}
		The corresponding effective phonon equation is then
		\begin{equation}
			\Big[
			D_0(\mathbf{k})
			-\omega^2 I
			+
			\Sigma_{\mathrm{ph}}(\mathbf{k},\omega)
			\Big]
			v(\mathbf{k},\omega)=0.
		\end{equation}
		
		In a time-reversal-broken magnetic background, the magnetic response contains an odd-in-frequency contribution proportional to the time-reversal-odd polarization $M_z$. At low frequency, this may be parameterized as
		\begin{equation}
			\chi_m^{\mathrm{odd}}(\mathbf{k},\omega)
			\sim
			i\omega\,
			\frac{M_z}{\Omega_m}\,
			\Gamma_m(\mathbf{k}),
		\end{equation}
		where $\Omega_m$ is the characteristic energy scale of the gapped magnetic fluctuation and $\Gamma_m(\mathbf{k})$ is a dimensionless matrix fixed by the magnetic dynamics and symmetry. Substituting this into the phonon self-energy gives the odd-in-frequency correction
		\begin{equation}
			\Sigma_{\mathrm{ph}}^{\mathrm{odd}}(\mathbf{k},\omega)
			\sim
			i\omega\,
			\lambda^2
			\frac{M_z}{\Omega_m}
			\Big[
			B(\mathbf{k})\,\Gamma_m(\mathbf{k})\,B^\dagger(\mathbf{k})
			\Big].
		\end{equation}
		
		The quantity in square brackets is naturally identified with a dimensionless bond-resolved kernel, which we denote by $\Gamma(\mathbf{k})$. The effective phonon equation may therefore be written as
		\begin{equation}
			\Big[
			D_0(\mathbf{k})-\omega^2 I+\omega G(\mathbf{k})
			\Big]v(\mathbf{k},\omega)=0,
		\end{equation}
		with $G(\mathbf{k})=g\,\Gamma(\mathbf{k})$ and $g \sim \lambda^2 M_z/\Omega_m$.
		
		We emphasize that this relation should be understood as a low-energy scaling estimate rather than as a fully microscopic equality with a universal prefactor. The detailed coefficient, and even the precise normalization of $\Omega_m$, depend on how the magnetic fluctuation and the magnetoelastic coupling are defined in a more microscopic theory. For the purposes of the present effective phonon model, however, the above form captures the essential physics.
	
	\section{Construction of $D_0(\mathbf k)$ and a time-odd kernel on the square-octagon lattice}
	
	We adopt the same Bloch convention as in the main text,
	\begin{equation}
		\label{eq:S_bloch}
		\mathbf u_{R,i}(t)
		=
		\boldsymbol{\epsilon}_i(\mathbf k)\,
		e^{i\mathbf k\cdot(R+\tau_i)}e^{-i\omega t}
	\end{equation}
	where $\mathbf r_{R,i}=R+\tau_i$ is the equilibrium position of basis site $i$ in cell $R$. It is convenient to collect the unnormalized Bloch amplitudes into the $N=ds$ component vector
	$\mathbf u(\mathbf k)=\big(\boldsymbol{\epsilon}_1(\mathbf k),\dots,\boldsymbol{\epsilon}_s(\mathbf k)\big)^T$. For a bond connecting $(R,i)$ to $(R+R_b,j)$, the displacement difference becomes
	
	\begin{equation}
		\label{eq:S_udiff}
		\mathbf u_{R+R_b,j}(t)-\mathbf u_{R,i}(t)
		=
		e^{i\mathbf k\cdot(R+\tau_i)}
		\Big(
		e^{i\mathbf k\cdot(R_b+\tau_j-\tau_i)}
		\boldsymbol{\epsilon}_j(\mathbf k)
		-
		\boldsymbol{\epsilon}_i(\mathbf k)
		\Big)e^{-i\omega t}
	\end{equation}
	
	Thus the off-diagonal coupling associated with this bond carries the phase factor
	$e^{i\mathbf k\cdot\mathbf d_{ij}^{(b)}}$, with
	$\mathbf d_{ij}^{(b)}\equiv R_b+\tau_j-\tau_i$. For a central spring on bond $b=(i\to j,R_b)$ with stiffness $K_b$ and bond vector $\mathbf d_{ij}^{(b)}$, we define the unit vector
	$\hat e_{ij}^{(b)}=\mathbf d_{ij}^{(b)}/|\mathbf d_{ij}^{(b)}|$. The harmonic bond energy is
	\begin{equation}
		\label{eq:S_Vbond}
		V_b
		=
		\frac{K_b}{2}
		\Big[
		\hat e_{ij}^{(b)}\cdot
		\big(
		\mathbf u_{R+R_b,j}-\mathbf u_{R,i}
		\big)
		\Big]^2
	\end{equation}
	which gives the force-constant block
	$\Phi_{ij}^{(b)}=K_b\,(\hat e_{ij}^{(b)}\otimes \hat e_{ij}^{(b)})$. In the Bloch basis, a single bond contributes an onsite restoring term to both connected sites and an off-diagonal hopping-like term between them. Summing over bonds gives the stiffness matrix blocks
	\begin{align}
		\label{eq:S_Kblocks}
		K_{ii}(\mathbf k)
		&=
		\sum_{b:\, i\leftrightarrow j}\Phi_{ij}^{(b)},
		\nonumber\\
		K_{ij}(\mathbf k)
		&=
		-\sum_{b:\, i\to j}
		\Phi_{ij}^{(b)}\,
		e^{i\mathbf k\cdot \mathbf d_{ij}^{(b)}}
		\qquad (i\neq j)
	\end{align}
	These expressions are the bond-resolved form of the Fourier transform
	\begin{align}
		\label{eq:S_KA_fourier}
		K_{ij}(\mathbf k)
		&=
		\sum_R
		\mathcal K_{ij}(R)\,
		e^{-i\mathbf k\cdot R}\,
		e^{i\mathbf k\cdot(\tau_j-\tau_i)}
		\nonumber\\
		A_{ij}(\mathbf k)
		&=
		\sum_R
		\mathcal A_{ij}(R)\,
		e^{-i\mathbf k\cdot R}\,
		e^{i\mathbf k\cdot(\tau_j-\tau_i)}
	\end{align}
	consistent with the Bloch phase convention in Eq.~\eqref{eq:S_bloch}.
	
	\noindent The mass matrix is block diagonal, $M=\mathrm{diag}(m_1 I_d,\,m_2 I_d,\,\dots,\,m_s I_d)$, and the mass-normalized conservative dynamical matrix is
	$D_0(\mathbf k)=M^{-1/2}K(\mathbf k)M^{-1/2}$. Equivalently, each bond $b=(i\to j,R_b)$ contributes
	\begin{align}
		\label{eq:S_D0bond}
		[D_0]_{ii}
		&\mathrel{+}=
		\frac{1}{m_i}\,\Phi_{ij}^{(b)}
		\qquad
		[D_0]_{jj}
		\mathrel{+}=
		\frac{1}{m_j}\,\Phi_{ij}^{(b)}
		\nonumber\\
		[D_0]_{ij}(\mathbf k)
		&\mathrel{+}=
		-\frac{1}{\sqrt{m_i m_j}}\,
		\Phi_{ij}^{(b)}\,
		e^{i\mathbf k\cdot\mathbf d_{ij}^{(b)}}
		\nonumber\\
		[D_0]_{ji}(\mathbf k)
		&=
		[D_0]_{ij}^{\dagger}(\mathbf k)
	\end{align}
		
	An effective time-odd velocity kernel adds a term linear in $\dot{\mathbf u}$ to the real-time equations of motion, and therefore contributes $-i\omega A(\mathbf k)$ in frequency space. In two dimensions, a standard antisymmetric structure is built from 
		$S_2= \begin{pmatrix}
			0 & -1\\
			1 & 0
		\end{pmatrix}$
	so that the real-space kernel $\mathcal A_{ij}(R)$ is real and antisymmetric. After Fourier transformation, $A(\mathbf k)$ is in general complex but satisfies the anti-Hermiticity required by the quadratic eigenvalue problem. Accordingly, after mass normalization, $\widetilde A(\mathbf k)=M^{-1/2}A(\mathbf k)M^{-1/2}$ is anti-Hermitian and $i\widetilde A(\mathbf k)$ is Hermitian. A common phenomenological scaling of the overall strength is $g \propto \frac{\lambda^2 M_z}{\Omega_m}$ where $\lambda$ denotes the coupling to a gapped internal mode of characteristic scale $\Omega_m$ in the presence of an out-of-plane time-reversal-odd polarization $M_z$.
	
	In mass-normalized variables,
	\begin{equation}
		\label{eq:S_QEP}
		\Big[
		D_0(\mathbf k)-\omega^2 I
		-i\omega \widetilde A(\mathbf k)
		\Big]\mathbf v(\mathbf k)=0
	\end{equation}
	
	where $\mathbf v(\mathbf k)=M^{1/2}\mathbf u(\mathbf k)$. It is often useful to introduce the Hermitian matrix $G(\mathbf k)\equiv -\,i\widetilde A(\mathbf k)$ so that Eq.~\eqref{eq:S_QEP} can be written as
	\begin{equation}
		\label{eq:S_QEP_G}
		\Big[
		D_0(\mathbf k)-\omega^2 I
		+\omega G(\mathbf k)
		\Big]\mathbf v(\mathbf k)=0
	\end{equation}
	Defining $\mathbf y=\omega \mathbf v$, we obtain the companion linearization
	\begin{equation}
		\label{eq:S_companion}
		\begin{pmatrix}
			0 & I\\
			D_0(\mathbf k) & G(\mathbf k)
		\end{pmatrix}
		\begin{pmatrix}
			\mathbf v\\
			\mathbf y
		\end{pmatrix}
		=
		\omega
		\begin{pmatrix}
			\mathbf v\\
			\mathbf y
		\end{pmatrix}
	\end{equation}
	
	Although both $D_0(\mathbf k)$ and $G(\mathbf k)$ are Hermitian, the full companion matrix in Eq.~\eqref{eq:S_companion} is generically non-Hermitian.
	
	\newpage
	\section{Phonon band structure in the absence of third neighbor coupling ($K_3=0$)}
	\label{app:phonon_bands_K3_0}
	
	\begin{figure}[!ht]
		\centering
		\includegraphics[width=\linewidth]{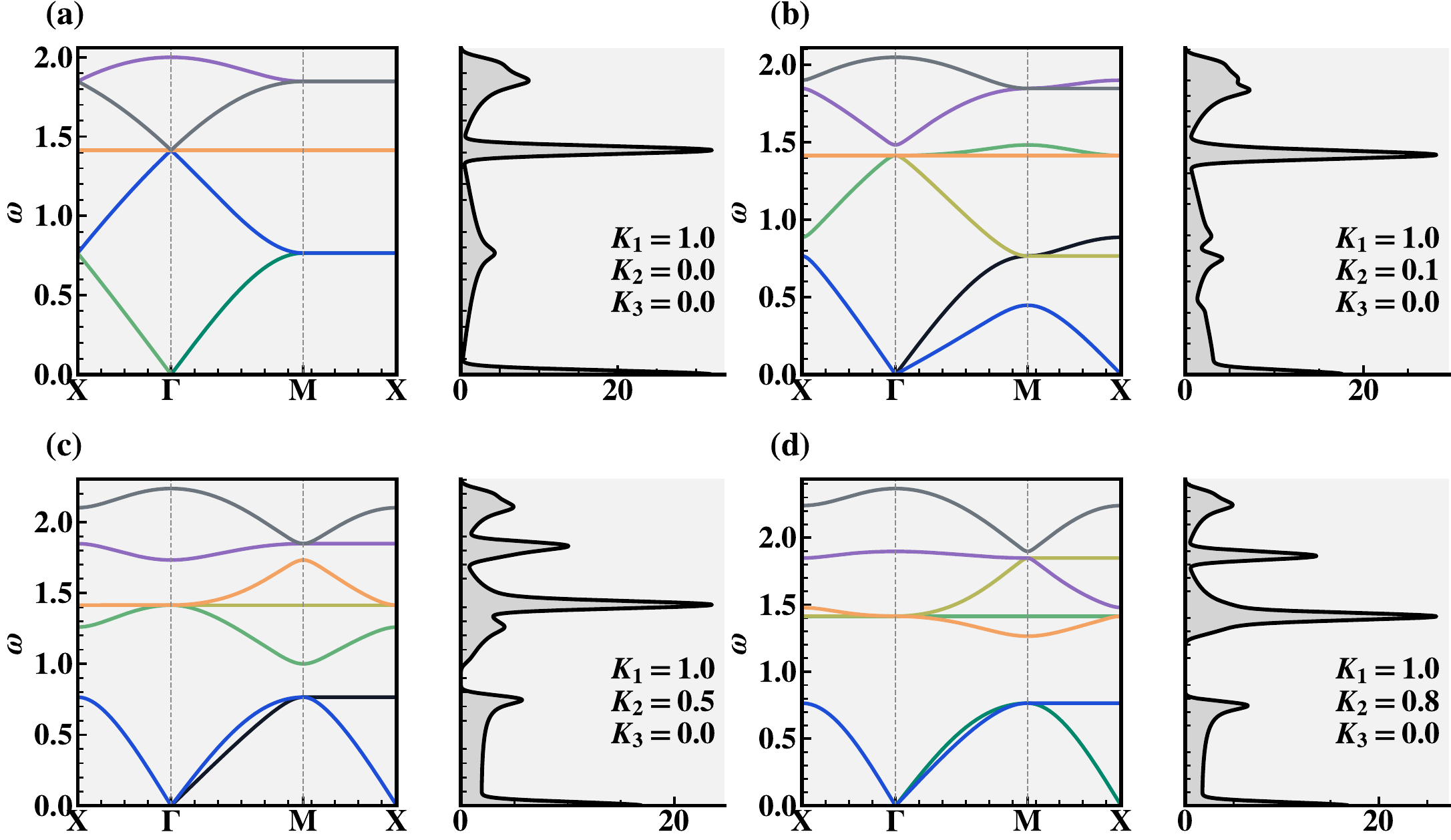}%
		\caption{
			Phonon dispersions along the $X$--$\Gamma$--$M$--$X$ path and the corresponding density of states for uniform masses and 3rd-NN coupling ($K_{3}=0$). The two acoustic branches are shown using darker colors(blue and black) to distinguish them from the optical sector.
		}
		\label{fig:K3_zero_uniform_masses}
	\end{figure}
	
\begin{figure}[!ht]
		\centering
		\includegraphics[width=\linewidth]{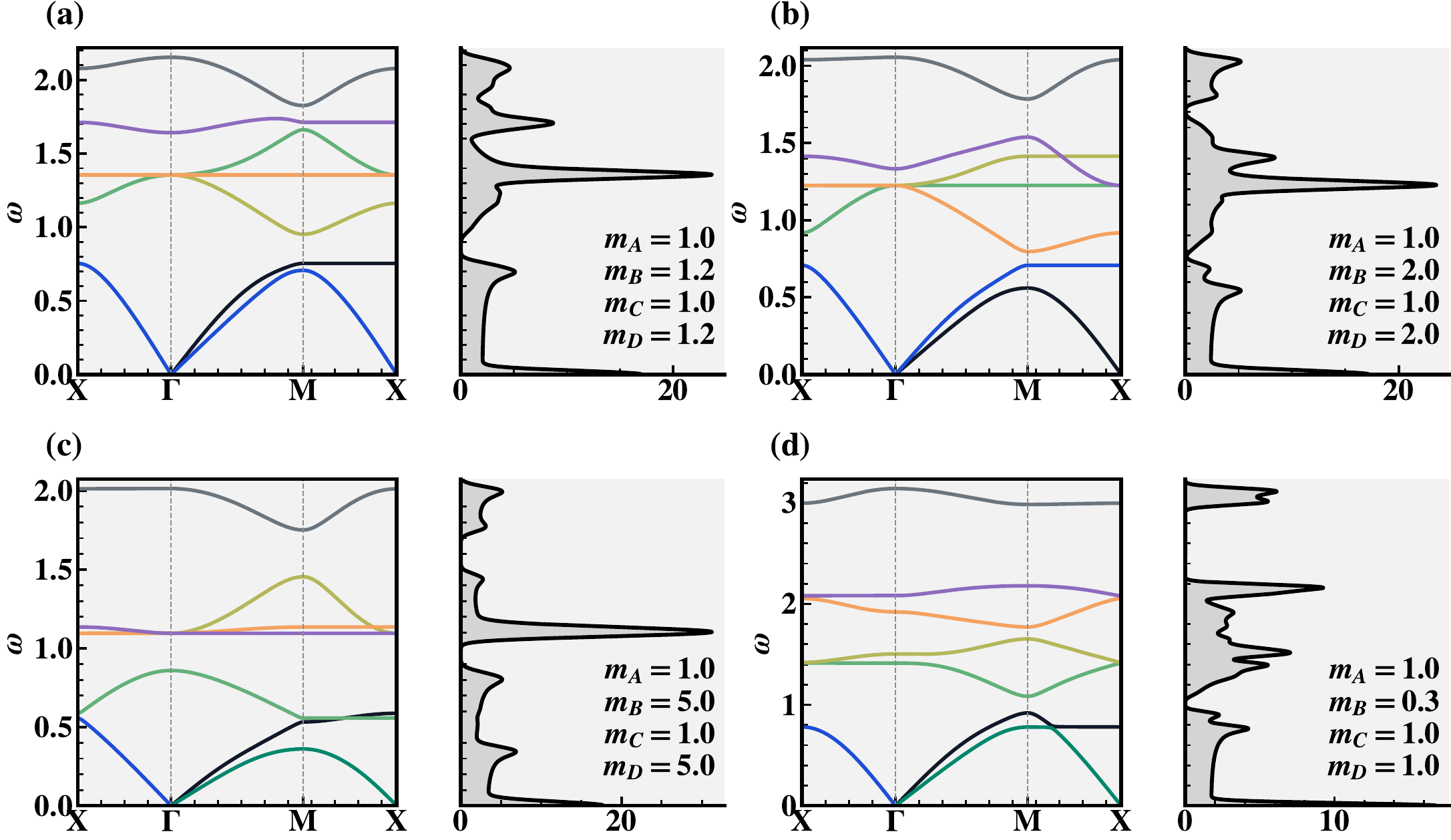}%
		\caption{
			Phonon dispersions along the $X$--$\Gamma$--$M$--$X$ path and the corresponding density of states for uniform masses and 3rd-NN coupling ($K_{3}=0$).
		}
		\label{fig:K3_zero_mass_variations}
	\end{figure}

	As a companion to the discussion in the main text for finite 3rd-NN coupling, here we discuss the phonon spectra of the square-octagon lattice in the limit $K_3=0$.  Figures~\ref{fig:K3_zero_uniform_masses} and \ref{fig:K3_zero_mass_variations} correspond to two instructive cases. In Fig.~\ref{fig:K3_zero_uniform_masses}, all masses are kept equal while the next-nearest-neighbor coupling $K_2$ is varied, so that the evolution of the dispersions reflects only the effect of longer-range restoring forces. In Fig.~\ref{fig:K3_zero_mass_variations}, by contrast, the spring constants are fixed and the mass pattern is varied, making it possible to isolate the influence of inertial imbalance on the band structure and density of states. A notable feature across all cases is the presence of two frozen modes, which produce sharp peaks in the DOS. For equal masses, these frozen modes occur at $q=0$ and $q=\sqrt{2}$.
	
	For uniform masses, increasing $K_2$ progressively reshapes the optical phonon bands. For the case of only NN allowed interaction (Fig.~\ref{fig:K3_zero_uniform_masses}(a)), several branches remain relatively flat over extended portions of the Brillouin-zone path. Once a weak next-nearest-neighbor coupling is introduced (Fig.~\ref{fig:K3_zero_uniform_masses}(b)), these nearly flat modes begin to disperse, and the corresponding DOS peaks broaden, signaling that the more localized vibrations of the nearest-neighbor problem are starting to hybridize with the extended lattice motion. At intermediate and larger $K_2$ (Figs.~\ref{fig:K3_zero_uniform_masses}(c) and (d)), this trend becomes more pronounced. The optical branches are reordered, their bandwidths change substantially, and the DOS weight is redistributed over a broader frequency range. 
	
	In contrast, the effect of mass variations is more qualitative. In Fig.~\ref{fig:K3_zero_mass_variations}, for $K_1 = 1.0$, $K_2 = 0.5$, and $K_3 = 0$, the introduction of mass imbalance rearranges the phonon modes themselves. For weak mass modulation (Fig.~\ref{fig:K3_zero_mass_variations}(a)), the spectrum is similar to the equal mass case, but the optical modes are already split. In Figs.~\ref{fig:K3_zero_mass_variations}(b) and (c), with increasing contrast, the heavier mass sites acquire lower frequency optical modes, and other modes are displaced, creating greater disparity among the modes. In the case of light mass impurity (Fig.~\ref{fig:K3_zero_mass_variations}(d)), the bandwidth is expanded and the spectrum is shifted upward, reflecting the greater frequencies associated with the lighter masses.
	
	\section{Berry curvature and Chern numbers on a discrete Brillouin zone}
	\label{subsec:chern_method}
	
	To characterize the topology of the phonon bands, we compute Berry fluxes and Chern numbers on a discretized Brillouin zone (BZ) using the gauge-invariant Fukui-Hatsugai-Suzuki (FHS) construction \cite{fukui2005chern}. Since the phonon problem is formulated as a quadratic eigenvalue problem in the frequency, we solve it through the companion linearization discussed above. The resulting companion matrix is generically non-Hermitian \cite{bergholtz2021exceptional}, and therefore the natural geometric framework is biorthogonal.
	
	For each band $n$, we determine right and left eigenvectors of the companion matrix,
	\begin{align}
		\label{eq:S_right_left_eigs}
		C(\mathbf k)\,|R_n(\mathbf k)\rangle
		&=
		\omega_n(\mathbf k)\,|R_n(\mathbf k)\rangle
		\nonumber\\
		\langle L_n(\mathbf k)|\,C(\mathbf k)
		&=
		\omega_n(\mathbf k)\,\langle L_n(\mathbf k)|
	\end{align}
	Because the companion linearization doubles the number of degrees of freedom, the full eigenvectors live in an enlarged space. To define the Berry geometry of the physical phonon modes, we retain only the displacement components of these eigenvectors, which we denote by $|r_n(\mathbf k)\rangle$ and $\langle \ell_n(\mathbf k)|$. These displacement vectors are normalized locally by the biorthogonality condition $\langle \ell_n(\mathbf k)|r_n(\mathbf k)\rangle=1$. We discretize the BZ on a periodic $N_k\times N_k$ mesh, $\mathbf k_{ij}=u_i\mathbf b_1+v_j\mathbf b_2$, where $\mathbf b_{1,2}$ are the reciprocal primitive vectors and $(u_i,v_j)$ are uniformly spaced in $[0,1)$. Let $\Delta \mathbf k_1$ and $\Delta \mathbf k_2$ denote one mesh step along the $\mathbf b_1$ and $\mathbf b_2$ directions, respectively.
	
	For an isolated band, we define the biorthogonal link variables
	\begin{equation}
		\label{eq:S_link_single}
		U_{\mu}(\mathbf k)=
		\frac{
			\langle \ell_n(\mathbf k)\,|\,
			r_n(\mathbf k+\Delta\mathbf k_{\mu})\rangle
		}{
			\big|
			\langle \ell_n(\mathbf k)\,|\,
			r_n(\mathbf k+\Delta\mathbf k_{\mu})\rangle
			\big|
		}
		\qquad
		\mu\in\{1,2\}
	\end{equation}
	These are the lattice analog of parallel-transport phases. The Berry flux through a plaquette based at $\mathbf k$ is then
	\begin{equation}
		\label{eq:S_plaquette_flux}
		F_{12}(\mathbf k)=
		\arg\!\Big[
		U_1(\mathbf k)\,
		U_2(\mathbf k+\Delta\mathbf k_1)\,
		U_1(\mathbf k+\Delta\mathbf k_2)^{-1}\,
		U_2(\mathbf k)^{-1}
		\Big]
	\end{equation}
	The corresponding Chern number is obtained from the BZ sum $C_n=\frac{1}{2\pi}\sum_{\mathbf k\in \mathrm{mesh}} F_{12}(\mathbf k)$. For a sufficiently fine mesh and an isolated band, this expression converges to an integer. A local Berry-curvature density may be approximated from the plaquette flux as
	\begin{equation}
		\label{eq:S_curv_density}
		\Omega_{12}(\mathbf k)\approx
		\frac{F_{12}(\mathbf k)}
		{\Delta k_1\,\Delta k_2}
	\end{equation}
	where $\Delta k_{1,2}$ are the mesh spacings along the two reciprocal directions.
	
	When two or more bands are nearly degenerate, the single-band construction can become numerically unstable, and it is more reliable to compute the Chern number of the corresponding multi-band subspace using the non-Abelian version of the FHS method \cite{fukui2005chern}. Let $R(\mathbf k)$ and $L(\mathbf k)$ be the matrices whose columns are the chosen right and left displacement vectors spanning the subspace at $\mathbf k$. We impose the subspace biorthonormality condition $L^\dagger(\mathbf k)\,R(\mathbf k)=I$. The overlap matrices between neighboring points are then
	\begin{equation}
		\label{eq:S_overlap_matrix}
		M_{\mu}(\mathbf k)=
		L^\dagger(\mathbf k)\,
		R(\mathbf k+\Delta\mathbf k_{\mu})
		\qquad
		\mu\in\{1,2\}
	\end{equation}
	from which we define the determinant link variables
	\begin{equation}
		\label{eq:S_link_subspace}
		U_{\mu}(\mathbf k)=
		\frac{\det M_{\mu}(\mathbf k)}
		{|\det M_{\mu}(\mathbf k)|}
	\end{equation}
	
	In practice, we additionally use overlap-based band tracking across the mesh in order to maintain consistent labeling of individual bands or multi-band subspaces. Although the FHS construction is gauge invariant, such tracking is useful when neighboring eigenvalues become close and a naive sorting by energy can interchange band indices.
	
	\bibliography{supp_references}